%% file: main-compsec.tex
\documentclass[5p]{elsarticle}

\PassOptionsToPackage{hyphens}{url}
\usepackage{lineno,hyperref}
\modulolinenumbers[5]

\journal{Computers \& Security}

\usepackage{graphicx}
\usepackage{verbatim}
\usepackage{amsmath}
\usepackage{tabularx}
\usepackage{multirow}
\usepackage{hhline}
\usepackage{color}
\usepackage{float}

\usepackage[nolist]{acronym}
\input{acronym.tex}

\usepackage{algpseudocode}
\usepackage{varwidth}
\usepackage{subcaption}

\usepackage{picins}

\usepackage{pifont}

\usepackage{threeparttable}

\usepackage{tikz}

\usepackage{cleveref}
\usepackage[ruled, vlined, longend, linesnumbered]{algorithm2e}
\SetAlFnt{\footnotesize}

\crefname{algocf}{Algorithm}{Algorithms}
\Crefname{algocf}{Algorithm}{Algorithms}

\crefname{section}{Sec.}{Secs.}
\Crefname{section}{Sec.}{Secs.}

\crefname{figure}{Fig.}{Figs.}
\Crefname{figure}{Fig.}{Figs.}

\usepackage{amsthm}

\newtheorem{lemma}{Lemma}
\newtheorem{theorem}{Theorem}

\begin{document}

\begin{frontmatter}
	
	\makeatletter
	\def\@author#1{\g@addto@macro\elsauthors{\normalsize%
			\def\baselinestretch{1}%
			\upshape\authorsep#1\unskip\textsuperscript{%
				\ifx\@fnmark\@empty\else\unskip\sep\@fnmark\let\sep=,\fi
				\ifx\@corref\@empty\else\unskip\sep\@corref\let\sep=,\fi
			}%
			\def\authorsep{\space and\space}%
			\global\let\@fnmark\@empty
			\global\let\@corref\@empty
			\global\let\sep\@empty}%
		\@eadauthor={#1}
	}
	\makeatother

\title{Accountable, Scalable and DoS-resilient Secure Vehicular Communication}

\author{Hongyu Jin\corref{cor1}}
\ead{hongyuj@kth.se}
\cortext[cor1]{Corresponding author}
\author{Panos Papadimitratos}
\ead{papadim@kth.se}

\address{Networked Systems Security Group\\KTH Royal Institute of Technology\\Stockholm, Sweden\\\url{https://www.eecs.kth.se/nss}}

\begin{abstract}
	Standardized \ac{VC}, mainly \acp{CAM} and \acp{DENM}, is paramount to vehicle safety, carrying vehicle status information and reports of traffic/road-related events respectively.
	Broadcasted \acp{CAM} and \acp{DENM} are pseudonymously authenticated for security and privacy protection, with each node needing to have all incoming messages validated within an expiration deadline.
	This creates an asymmetry that can be easily exploited by external adversaries to launch a clogging \ac{DoS} attack: each forged \ac{VC} message forces all neighboring nodes to cryptographically validate it; at increasing rates, easy to generate forged messages gradually exhaust processing resources and severely degrade or deny timely validation of benign \acp{CAM}/\acp{DENM}.
	The result can be  catastrophic when awareness of neighbor vehicle positions or critical reports are missed.
	We address this problem making the standardized \ac{VC} pseudonymous authentication~\cite{ieee160912,etsicam,etsidenm} \emph{\ac{DoS}-resilient}. We propose efficient cryptographic constructs, which we term message verification \emph{facilitators}, to prioritize processing resources for verification of potentially valid messages among bogus messages and verify multiple messages based on one signature verification.
	Any message acceptance is strictly based on public-key based message authentication/verification for \emph{accountability}, i.e., \emph{non-repudiation} is not sacrificed, unlike symmetric key based approaches.
	This further enables drastic \emph{misbehavior detection}, also exploiting the newly introduced facilitators, based on probabilistic signature verification and cross-checking over multiple facilitators verifying the same message; while maintaining verification latency low even when under attack, trading off modest communication overhead.
	Our facilitators can also be used for efficient discovery and verification of \emph{\ac{DENM}} or any \emph{event-driven message}, including \emph{misbehavior evidence} used for our scheme.
	Even when vehicles are saturated by adversaries mounting a clogging DoS attack,  transmitting high-rate bogus \acp{CAM}/\acp{DENM}, our scheme achieves an average $50 ms$ verification delay with message expiration ratio less than 1\% - a huge improvement over the current standard that verifies every message signature in a \ac{FCFS} manner and suffers from having 50\% to nearly 100\% of the received benign messages expiring.

\end{abstract}

\begin{keyword}
	Accountability\sep Non-repudiation\sep Privacy\sep Pseudonymous authentication\sep Efficiency
\end{keyword}

\end{frontmatter}

\acresetall

\input{section_compsec/introduction}
\input{section_compsec/related}
\input{section_compsec/problem}
\input{section_compsec/scheme}
\input{section_compsec/analysis}
\input{section_compsec/simulation}
\input{section_compsec/conclusion}

\section*{Acknowledgement}

This work was supported in parts by the Swedish Research Council and the Knut och Alice Wallenberg Foundation.

\bibliography{references}


\end{document}

%% file: acronym.tex
\begin{acronym}
	\acro{AAA}{Authentication, Authorization and Accounting}
	\acro{AP}{Access Point}
	\acro{API}{Application Programming Interface}
	\acro{ASIC}{Application-Specific Integrated Circuit}
	\acro{BP}{Baseline Pseudonym}
	\acro{BS}{Base Station} 
    \acro{BSM}{Basic Safety Message}
    \acro{BYOD}{Bring Your Own Device}
   	\acro{C-V2X}{Cellular-V2X}
	\acro{C2C-CC}{Car2Car Communication Consortium}
	\acro{CA}{Certification Authority}
	\acrodefplural{CA}{Certification Authorities}
	\acro{CN}{Common Name}
	\acro{CAM}{Cooperative Awareness Message}
	\acro{CDF}{Cumulative Distribution Function}
	\acro{CIA}{Confidentiality, Integrity and Availability}
	\acro{CRL}{Certificate Revocation List}
	\acro{CDN}{Content Delivery Network}
	\acro{CSR}{Certificate Signing Requests}
	\acro{DAA}{Direct Anonymous Attestation}
	\acro{DDoS}{Distributed Denial of Service}
	\acro{DDH}{Decisional Diffie-Helman}
	\acro{DENM}{Decentralized Environmental Notification Message}
	\acro{DHT}{Distributed Hash Table}
	\acro{DoS}{Denial of Service}
	\acro{DPA}{Data Protection Agency}
	\acro{DSRC}{Dedicated Short Range Communication}
	\acro{EC}{Elliptic Curve}
	\acro{ECC}{Elliptic Curve Cryptography}
	\acro{ECDSA}{Elliptic Curve Digital Signature Algorithm}
	\acro{ECU}{Electronic Control Unit}
	\acro{ETSI}{European Telecommunications Standards Institute}
	\acro{EVITA}{E-safety Vehicle Intrusion protected Applications}
	\acro{FCFS}{First-Come First-Served}
	\acro{FOT}{Field Operational Test}
	\acro{FPGA}{Field-Programmable Gate Array}
	\acro{GN}{GeoNetworking}
	\acro{GNSS}{Global Navigation Satellite System}
	\acro{GS}{Group Signatures}
	\acro{GM}{Group Manager}
	\acro{GBA}{Generic Bootstrapping Architecture}
	\acro{GPA}{Global Passive Adversary}
	\acro{GUI}{Graphic User Interface}
	\acro{HMAC}{Hash-based Message Authentication Code}
	\acro{HP}{Hybrid Pseudonym}
	\acro{HSM}{Hardware Security Module}
	\acro{HTTP}{Hypertext Transfer Protocol}
	\acro{IEEE}{Institute of Electrical and Electronics Engineers}
	\acro{IoT}{Internet of Things}
	\acro{ITS}{Intelligent Transport System}
	\acro{IT}{Information Technologies}
	\acro{IMSI}{International Mobile Subscriber Identity}
	\acro{IMEI}{International Mobile Station Equipment Identity}
	\acro{IdP}{Identity Provider}
	\acro{ISP}{Internet Service Provider}
	\acro{KRL}{Key Chain Revocation List}
	\acro{LCFS}{Last-Come-First-Served}
	\acro{LEA}{Law Enforcement Agency}
	\acro{LTC}{Long-Term Certificate}
	\acro{LTCA}{Long-Term \acl{CA}}
	\acrodefplural{LTCA}{Long-Term \aclp{CA}}
	\acro{LDAP}{Lightweight Directory Access Protocol}
	\acro{LTE}{Long Term Evolution}
	\acro{LBS}{Location-based Service}

	\acro{MAC}{Message Authentication Code}
	\acro{MCA}{Message \ac{CA}}
	\acro{MEA}{Misbehavior Evaluation Authority}
	\acro{NTP}{Network Time Protocol}
	\acro{OBU}{On-Board Unit}
	\acro{OCSP}{Online Certificate Status Protocol}
	\acro{OSN}{Online Social Network}
	\acro{PC}{Pseudonymous Certificate}
	\acro{PCA}{Pseudonymous \acl{CA}}
	\acrodefplural{PCA}{Pseudonymous \aclp{CA}}
	\acro{PDP}{Policy Decision Point}
	\acro{PEP}{Policy Enforcement Point}
	\acro{PIR}{Private Information Retrieval}
	
	\acro{PKI}{Public-Key Infrastructure}
	\acro{POI}{Point of Interest}
	\acro{PRECIOSA}{Privacy Enabled Capability in Co-operative Systems and Safety Applications}
	\acro{PRESERVE}{Preparing Secure Vehicle-to-X Communication Systems}
	\acro{PRL}{Pseudonymous Certificate Revocation List}
	\acro{P2P}{peer-to-peer}
	\acro{RA}{Resolution Authority}
	\acro{REST}{Representational State Transfer}
	\acro{RBAC}{Role Based Access Control}
	\acro{RCA}{Root \acl{CA}}
	\acro{RSU}{Roadside Unit}
	\acro{SAML}{Security Assertion Markup Language}
	\acro{SCORE@F}{Système COopératif Routier Expérimental Français}
	\acro{SeVeCom}{Secure Vehicle Communication}
	\acro{SIS}{Security Infrastructure Server}
	\acro{SP}{Service Provider}
	\acro{SSO}{Single-Sign-On}
	\acro{SoA}{Service-oriented-Approach}
	\acro{SOAP}{Simple Object Access Protocol}
	\acro{SAS}{Sample Aggregation Service}
	\acro{TESLA}{Timed Efficient Stream Loss-Tolerant Authentication}
	\acro{TS}{Task Service}
	\acro{TLS}{Transport Layer Security}
	\acro{TPM}{Trusted Platform Module}
	\acro{TTP}{Trusted Third Party}
	\acro{TVR}{Ticket Validation Repository}
	\acro{URI}{Uniform Resource Identifier}
	\acro{VANET}{Vehicular Ad-hoc Network}
	\acro{V2I}{Vehicle-to-Infrastructure}
	\acro{V2V}{Vehicle-to-Vehicle}
	\acro{V2X}{Vehicle-to-Everything}
	\acro{VC}{Vehicular Communication}
	\acro{VM}{Virtual Machine}
	\acro{VSN}{Vehicular Social Network}
	\acro{VSS}{\ac{VC} Security Subsystem}
	\acro{WAVE}{Wireless Access in Vehicular Environments}
	\acro{WSDL}{Web Services Discovery Language}
	\acro{W3C}{World Wide Web Consortium}
	\acro{VANET}{Vehicular Ad-hoc Network}
	\acro{VPKI}{Vehicular Public-Key Infrastructure}
	\acro{WS}{Web Service}
	\acro{WoT}{Web of Trust}
	\acro{WSACA}{\ac{WAVE} Service Advertisement \ac{CA}}
	\acro{XML}{Extensible Markup Language}
	\acro{XACML}{eXtensible Access Control Markup Language}
	\acro{3G}{3rd Generation}
\end{acronym}

%% file: section_compsec/introduction.tex
\section{Introduction}
\label{sec:introduction}

Standardized \acp{CAM} and \acp{DENM} enable \ac{V2V} data exchange~\cite{ieee160912,etsicam,etsidenm}.
\acp{CAM}, or \emph{safety beacons}, are broadcasted at a rate from 1$Hz$ to 10$Hz$, allowing each vehicle to maintain a view of neighboring vehicle mobility. 
Event-driven \acp{DENM} inform about, for example, abnormal or hazardous road situations.
Standards mandate that \acp{CAM} and \acp{DENM} are secured, in particular pseudonymously authenticated, for security and message unlinkability across distinct pseudonym lifetime periods.
More specifically, they are digitally signed (\ac{ECDSA}) with each vehicle holding short-lived private/public keys and anoynimized certificates, termed pseudonyms or \acp{PC}~\cite{papadimitratos2008secure,kargl2008secure,papadimitratos2009vehicular,khodaei2014towards,khodaei2015key,khodaei2018secmace}.
\ac{ECDSA} is used instead of the more popular Internet/Web RSA algorithm, because \ac{EC} key sizes and \ac{ECDSA} signatures are much smaller than RSA ones for the same security level~\cite{ieee16092} (at the expense of higher signature verification delay).

High rate communication, often, dense \ac{VC} network neighborhoods and frequently changing \acp{PC}, for improved unlinkability, create an inherent asymmetry: each sent message must be validated by all neighbors and, at each vehicle, incoming messages need to be validated within a short time and have an expiration deadline.
ECDSA signature verification delay, $\tau$, is in the order of milliseconds~\cite{calandriello2011performance,petit2013authentication,baee2019broadcast,pan2019secure,preserved32,campvsc5,mehrabi2022efficient}.
Even in moderately dense network settings, with a $\tau=4\ msec$, an \ac{OBU} can verify up to 250 messages per second; that is, for a typical rate of 10 \acp{CAM}/second transmitted by each OBU, from 25 neighboring vehicles.
The larger/denser the neighborhood, the higher the traffic; around 2000 \acp{CAM}/second could be sent for the 802.11p default data rate of 6 Mbps~\cite{sepulcre20176} (\acp{CAM} of 300 bytes~\cite{c2c-camstat}), an order of magnitude more messages than what a typical \ac{OBU} can cryptographically handle.

\begin{figure}[t]
	\centering
	\includegraphics[width=\linewidth]{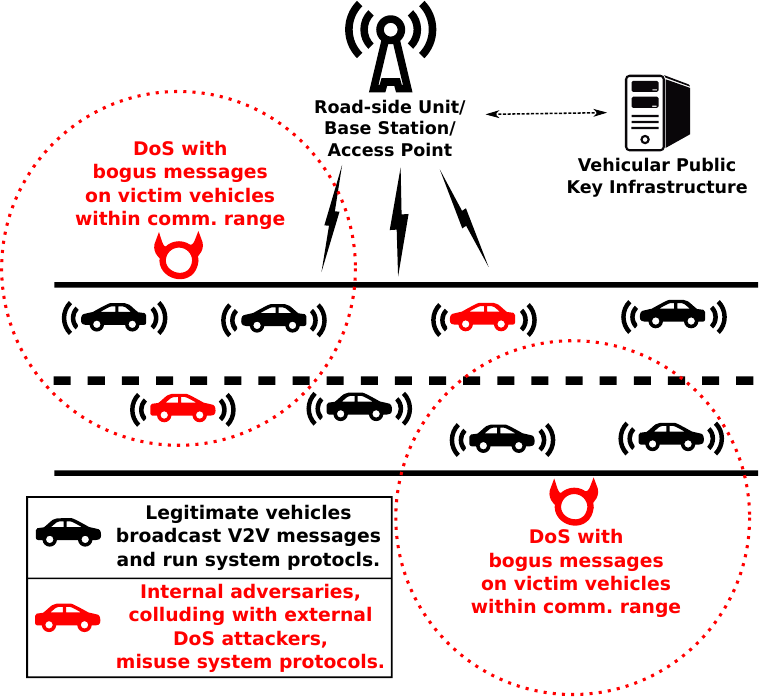}
	\caption{\acf{VC} system under \ac{DoS} attacks.}
	\label{fig:system}
\end{figure}

Adversaries can exploit this asymmetry by transmitting fast to generate bogus messages, with bogus signatures and \acp{PC}, forcing their receiving neighbors to validate them all, as illustrated in~\cref{fig:system}\footnote{We explain in \cref{sec:problem} the internal adversary.}.
Receivers will reject bogus messages, but validating one or two signatures per bogus message delays benign message processing, easily beyond the expiration deadline.
Such clogging \ac{DoS} attacks can target and affect both \acp{CAM} and \acp{DENM} (or any secure event-driven messages); high verification latencies and high expiration ratios, especially for highly critical \acp{DENM}, could be fatal.

\textbf{Challenges.} Handling security overhead in VC, important even without clogging \ac{DoS}, has received attention: optimizations~\cite{calandriello2007efficient,schoch2010efficiency,calandriello2011performance,feiri2014formal,jin2016proactive} reduce beacon size and skip \ac{PC} validation if already verified and cached locally; with limited or no effect for high-rate beacon arrivals imposing excessive computation overhead.
Combining public key and symmetric key based beacon authentication reduces overhead~\cite{studer2009flexible,hsiao2011flooding,lyu2016pba}.
But such an approach has two shortcomings: it cannot provide an efficient way to discover that first beacon (requiring the sender's signature and its PC verification) to `bootstrap' subsequent symmetric key beacon verification.
Moreover, it does not provide non-repudiation and accountability for symmetric-key-authenticated messages (see the explanation in \cref{sec:related} with \cref{fig:bogus}); both mandatory for \ac{VC} security requirements~\cite{papadimitratos2006securing,etsisecurity} (\emph{Challenge 1}).

Cooperating vehicles can share their verification efforts in order to accelerate queue processing~\cite{lin2013achieving,jin2015scaling}, but fail to protect vehicles from clogging \ac{DoS} attacks.
Moreover, there is no efficient way to detect misbehavior sharing false verification results (\emph{Challenge 2}).
The challenge lies exactly in that adversaries can transmit messages with bogus signatures and PCs forcing their receiving neighbors to verify signatures before dropping the bogus messages.
What is lacking is an efficient approach to filter out bogus messages without verifying every message signature while preserving non-repudiation and accountability.

Last but not least, in spite the aforementioned works to render beacon verification efficient~\cite{calandriello2007efficient,schoch2010efficiency,calandriello2011performance,feiri2014formal,studer2009flexible,hsiao2011flooding,lyu2016pba,lin2013achieving,jin2015scaling,jin2016proactive}, the literature has not considered \ac{DoS} attacks targeting \textit{event-driven messages} (or event messages, for simplicity) (\emph{Challenge 3}).
Unlike periodic high-frequency safety beacons, each event message (e.g., \acp{DENM}) could be independent from earlier messages.
It can also carry information of varying vehicle safety criticality.
Thus strict signature verification is necessary for event messages, without alternative efficient verification approaches (applicable for safety beacons).
Simply put, high verification latencies and high expiration ratios caused by \ac{DoS} attacks could be fatal.

\begin{figure}[t]
	\centering
	\includegraphics[width=\linewidth]{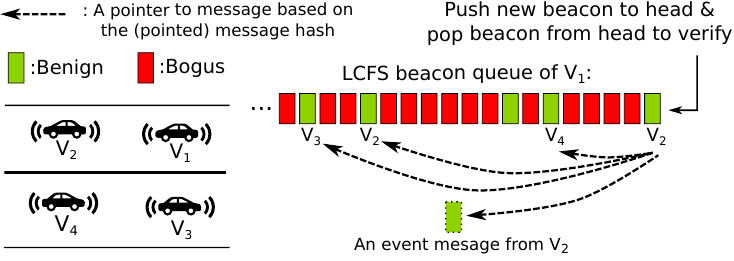}
	\caption{Illustration of cooperative verification under a \ac{DoS} attack.}
	\label{fig:intro}
\end{figure}

\textbf{Contributions.} Concerned with cryptographic message verification\footnote{Message content validation is out of the scope of this paper; addressed by approaches such as~\cite{raya2008data,gisdakis2015shield} orthogonal to our scheme.}, we address the three aforementioned challenges: we extend traditional, standard-compliant pseudonymous authentication for \ac{VC}, integrating cooperative beacon verification and symmetric-key constructs to accelerate message triage.
Our scheme encompasses/safeguards both \acp{CAM} (or beacons) and \acp{DENM} from clogging \ac{DoS}, and to the best of our knowledge, is the first to support \ac{DoS}-resilient \ac{V2V} \ac{DENM} validation.
In summary, our contributions here are:
\begin{enumerate}
	\item Non-repudiable DoS-resilient CAM and DENM authentication, strictly based on public key cryptography, assisted by resilient cooperative verification and neighbor vehicle discovery.
	\item Misbehavior detection with probabilistic cooperative verifier checking and verifier cross-checking.
	\item DoS-resilient event-driven message verification, including \ac{DENM} or any event-driven messages (e.g., misbehavior evidence used for our scheme).
\end{enumerate}

We extend safety beacons with extra fields (e.g., message verification facilitators) that enable verification of queued beacons and facilitate benign event message verification.
Intuitively, verifying one beacon signature can possibly validate more than one beacons, thus expediting message verification under high message arrival rate.
\Cref{fig:intro} shows an example of cooperative verification with a \ac{LCFS} queue processing (see \Cref{sec:scheme} for the explanation of the \ac{LCFS} design choice).
Vehicle $V_1$ pops a beacon from the queue head.
Once this beacon, sent by $V_2$, is verified, the beacon can (cooperatively) allow verifying a beacon from $V_3$ and a beacon form $V_4$.
It can also validate immediate earlier beacons from $V_2$, if not verified yet.
It also carries a facilitator that points to the upcoming event message, to help receiver discovering the corresponding benign event message among bogus event messages.
The event message was already triggered (and generated) before the beacon dissemination, but it was artificially delayed to facilitate the event message reception and verification.
Although our scheme delays the event message dissemination, still, it outperforms the traditional approach under \ac{DoS} attacks (see \Cref{sec:simulation}).

Hash chain elements are attached to beacons, to efficiently eliminate masqueraded beacons.
Once a neighboring vehicle is discovered (based on beacon reception and validation), the hash chains help keeping track of subsequent beacons from the already discovered neighboring vehicles, which still need to go through signature verification to achieve non-repudiation and accountability unlike prior approaches~\cite{studer2009flexible,lyu2016pba,jin2019resilient} (\emph{Contribution 1}).
When overloaded by high rate benign beacons or DoS attacks, vehicles can expedite the discovery of new neighboring vehicles leveraging piggybacked information on the (discovered) neighbors' beacons.
Moreover, probabilistic signature checking (of cooperatively verified beacons) and cross-checking of multiple validation elements effectively thwart malicious nodes that attempt to exploit the cooperative verification (\emph{Contribution 2}). The event message facilitators in \acp{CAM} help to efficiently discover the corresponding upcoming \acp{DENM} or any event-driven messages, including misbehavior evidence introduced for our scheme in \cref{sec:scheme} (\emph{Contribution 3}).

What we are after in this paper is a \ac{DoS}-resilient efficient beacon verification scheme that preserves non-repudiation.
Our scheme introduces extra communication overhead but provides timely beacon and event message validation under \ac{DoS} attacks; something impossible for standardized secure \ac{VC}.
Moreover, our approach is agnostic to the underlying communication technology as it is proposed based on the functional specification of the safety application.

In the rest of the paper, we discuss related works (\Cref{sec:related}) and explain the system and the adversarial models (\Cref{sec:problem}).
Then, we provide a detailed description of our scheme (\Cref{sec:scheme}).
We analyze the security properties fulfilled by our scheme (\Cref{sec:analysis}), we quantitatively evaluate the scalability and resilience to \ac{DoS} attacks and misbehaving malicious nodes, with simulations based on realistic vehicle mobility, communication module and processing delays (\Cref{sec:simulation}), before concluding (\Cref{sec:conclusion}).

%% file: section_compsec/related.tex
\section{Background and Related Works}
\label{sec:related}
	
\begin{table*}[t]
	\newcommand{\tabincell}[2]{\begin{tabular}{@{}#1@{}}#2\end{tabular}}
	\caption{Functional and security properties achieved by related works and our scheme.}
	\centering
	\small
	\begin{threeparttable}
	\begin{tabular}{| c | c | c | c | c | c | c |}
		\hline 
		Scheme & \tabincell{c}{Hardware \\ Flexibility\tnote{a}} & \tabincell{c}{CAM \\ Support} & \tabincell{c}{DENM \\ Support} & Non-repudiation & DoS-resilience & \tabincell{c}{Efficient \\ Neighbor \\ Discovery}  \\\hline
		\tabincell{c}{Hardware Acceleration \cite{mehrabi2022efficient}} & \ding{54} & \ding{52} & \ding{52} & \ding{52} & \ding{54} & \ding{54} \\\hline
		\tabincell{c}{PC Verification \\ Optimization~\cite{calandriello2011performance,jin2016proactive}} & \ding{52} & \ding{52} & \ding{52} & \ding{52} & \ding{54}  & \ding{54} \\\hline
		\tabincell{c}{Context-adaptive \\ Verification~\cite{schoch2010efficiency}} & \ding{52} & \ding{52} & \ding{54} & \ding{52} & \ding{54} & \ding{54} \\\hline
		\tabincell{c}{Symmetric Key-based \\ Authentication~\cite{hu2006strong,studer2009flexible,hsiao2011flooding, lyu2016pba}} & \ding{52} & \ding{52} & \ding{54} & \ding{54} & \ding{52} & \ding{54} \\\hline
		\tabincell{c}{Cooperative Verification~\cite{lin2013achieving,jin2015scaling}} & \ding{52} & \ding{52} & \ding{54} & \ding{52} & \ding{54} & \ding{54} \\\hline
		\tabincell{c}{DoS-resilient \\ Cooperative Verification~\cite{jin2019resilient}} & \ding{52} & \ding{52} & \ding{54} & \ding{54} & \ding{52} & \ding{52} \\\hline
		\tabincell{c}{Physical Layer \\ Fingerprinting~\cite{dongre2021message}} & \ding{54} & \ding{52} & \ding{52} & \ding{52} & \ding{52}\textsuperscript{{\kern-0.5em\ding{54}}} \tnote{b} & \ding{54} \\\hline
		\tabincell{c}{Puzzles for \\ V2I Communication~\cite{sun2017privacy,liu2018mitigating}} & \ding{52} & \ding{54} & \ding{54} & \ding{52} & \ding{52} & \ding{54} \\\hline
		Our Scheme & \ding{52} & \ding{52} & \ding{52} & \ding{52} & \ding{52} & \ding{52} \\
		\hline
	\end{tabular}
	\begin{tablenotes}
		\item[a] The hardware flexibility column indicates the solutions rely on special hardware (\ding{54}) or are flexible in required hardware (\ding{52}).
		\item[b] This solution relies on non-cryptographic-based bogus message filtering with high inaccuracy and there is no evidence on fingerprinting efficiently, compared to cryptographic signature verification.
	\end{tablenotes}
	\end{threeparttable}
	\label{table:related}
\end{table*}
	
A number of studies have focused on addressing cryptographic computational overhead in \ac{V2V} communication and mitigating \ac{DoS} attacks.
This section provides an overview of these related works, addressing the problem at hand with various viewpoints, with strengths and weaknesses in terms of functional and security properties we are after in this paper, summarized in \Cref{table:related}.

\textbf{Hardware Acceleration:} This can be a solution for handling cryptographic overhead in an energy efficient manner with relatively lower cryptographic processing delays.
However, flexibility is lacking: a hardware accelerator typically supports one or a few algorithms, and can fail in handling changes, in our context, in supporting new cryptographic primitives, e.g., new elliptic curves.
Moreover, it may not be realistic to expect that powerful yet costly hardware is universally available for all vehicles on the road.
On one hand, solutions available for both powerful or budget devices are necessary.
On the other hand, if powerful low cost cryptographic hardware were universally available, a security level increase would be necessary to counter brute force attacks, which would then lead to higher cryptographic processing delays~\cite{lenstra2001selecting,jin2024future}.
As a result, hardware acceleration would still be insufficient to counter clogging DoS.

Recent results regarding hardware acceleration indeed indicate a lack of resilience to clogging DoS attacks.
For example, a recent hardware accelerator for brainpoolP256r1 signature verification~\cite{mehrabi2022efficient} takes around 1 $ms$ with a standard double-and-add point multiplication~\cite{hankerson2006guide}.
It is only able to handle at most 1000 signature verification per second - much less than the 2000 messages per second the default 802.11p data rate could allow an adversary to transmit. One should additionally consider that expected performance in a real world deployment could be worse than the optimistic hardware benchmarks.
Our proposal thwarts clogging DoS without specific hardware performance requirements, remaining effective as on-board processing (and data rates and security levels) may increase in the future. 

\textbf{Security Overhead Optimization:} The baseline, the standard approach is that vehicles perform signature verification for each and every received message in a \ac{FCFS} manner.
Optimizations~\cite{calandriello2011performance,jin2016proactive} reduce the number of required signature verifications on \acp{PC}, but not for the \acp{CAM}/\acp{DENM}.
Context-adaptive beacon verification~\cite{schoch2010efficiency} processes message signatures probabilistically based on the context, minimizing the effect of bogus beacons, but it is not able to efficiently filter out beacons forged precisely to mislead that they were sent by newly encountered neighboring vehicles.

Neighboring vehicles can collaborate to verify received beacons in order to benefit each other with locally and independently performed signature verifications~\cite{lin2013achieving,jin2015scaling}.
Vehicles share own recent message validation results, leveraging their own beacon disseminations so that a signature verification can enable the receiving neighbor to verify more than one beacons in its local queue~\cite{jin2015scaling}.
However, cooperative verification requires an efficient approach to find out benign messages, among high rate bogus messages, to make use of shared verification.
Alternatively, vehicles can leverage \ac{RSU}-aided collaborative verification based on ID-based signcryption~\cite{lin2013achieving}; each vehicle verifies a subset of messages and is informed about the rest of the message validations from other vehicles.
However, processing delays of this \ac{RSU}-aided approach are not provided, making it unclear whether the approach could outperform traditional public-key cryptography based schemes.
Moreover, the approach considers verification of a set of given messages, but does not consider dynamic and continuous message reception and verification, as is the case for high-rate safety beacons (\acp{CAM}).
Last but not least, the scheme is not resilient to compromised vehicles.
Although the above optimizations relieve the computation overhead to a certain extent, essentially aiming at lower average message verification delay (namely $\tau$ in this paper), an overwhelming bogus beacon rate could force vehicles to dedicate/waste the majority of computation resource in verifying bogus signatures.
Our proposal addresses this challenge by efficiently discovering potentially valid benign messages and filtering out bogus messages.

\textbf{Symmetric Key based Authentication:} Prediction-based approaches rely on the predictability of vehicle status given practical mobility limitations~\cite{hsiao2011flooding, lyu2016pba}.
However, the prediction-based approaches are not resilient to packet loss, which is common in a loaded \ac{VANET}.
\ac{TESLA}~\cite{perrig2000efficient}-based schemes~\cite{hu2006strong,studer2009flexible} leverage one-way hash chains as symmetric keys to authenticate beacons while coping with packet loss~\cite{hu2006strong}.
However, symmetric key based authentication is possible only after at least one beacon is verified, so that the hash chain elements can be connected to the authenticated chain anchor.
Moreover, creating the asymmetry with the help of delayed symmetric key disclosure precludes non-repudiation and accountability because message authenticators can be forged once the symmetric keys are disclosed.
\ac{DoS}-resilient secure communication in wireless sensor networks~\cite{dong2013providing} has been proposed to defend against clogging \ac{DoS} attacks, discovering potentially valid messages leveraging hash chains.
However, the trust establishment on the hash chain in an dynamic network is not addressed, clearly needed in \ac{VC} systems.
Our scheme adopts and extends the \ac{DoS}-resilient message discovery~\cite{dong2013providing}, as a feature towards achieving \ac{DoS}-resilient \ac{VC} systems.

\begin{figure}[t]
	\centering
	\includegraphics[width=0.8\linewidth]{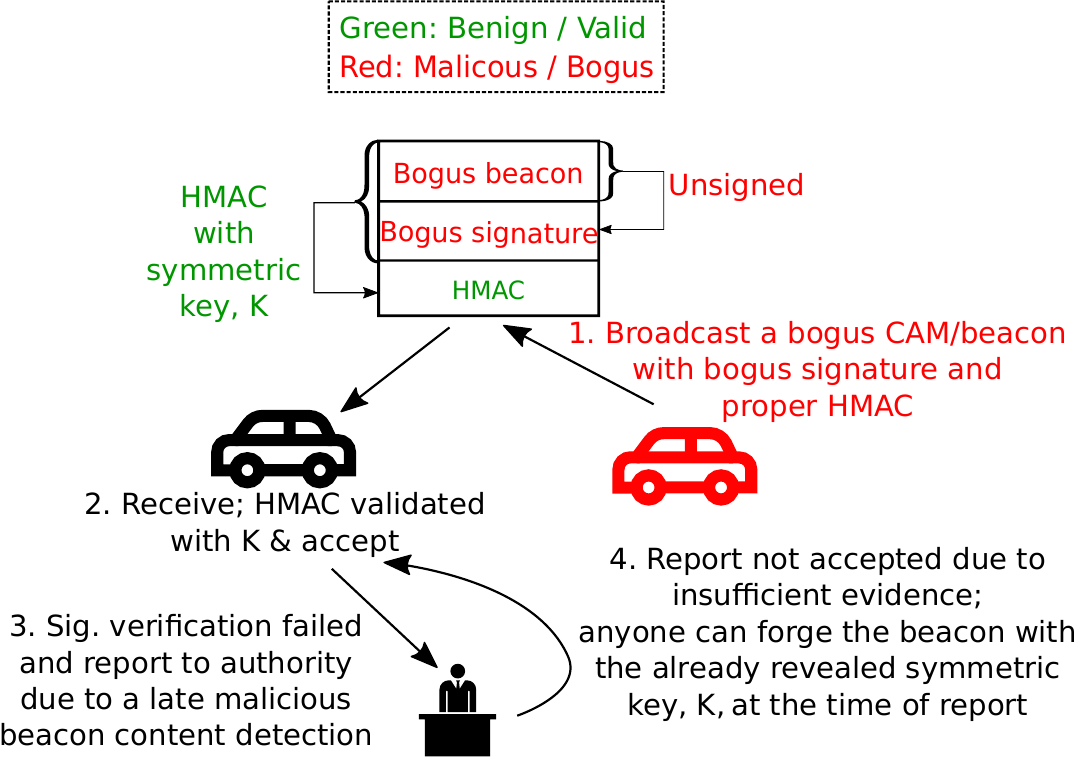}
	\caption{An attacker exploits the lack of non-repudiation in the symmetric key chain based solution.}
	\label{fig:bogus}
\end{figure}

To the best of our knowledge, only \cite{jin2019resilient} provides \ac{DoS}-resilient features for newly encountered vehicle discovery.
However, similar to~\cite{hu2006strong,studer2009flexible,lyu2016pba}, it relies on symmetric key based authentication for efficient beacon validation, which fails providing non-repudiation.
Beacons can be properly signed on top of symmetric key based authentication in order to protect against misusing the absence of non-repudiation~\cite{studer2009flexible,jin2019resilient}, while
receivers choose to verify message signatures if message non-repudiation is required.
However, it is hard to define conditions for signature verification, while late detection of misbehavior could result in benign vehicles being misled by considerable amount of fake beacons without making the malicious vehicles accountable for their actions as shown in~\cref{fig:bogus}.
Our scheme ensures non-repudiation of message authentication and accountability of vehicles by strictly accepting messages based on public key cryptography.

All the above approaches exploit high beacon frequency and correlation of successive beacons, but they cannot thwart \ac{DoS} attacks on event-driven message.
Event messages does not follow a specific time pattern, and are triggered by less correlated or independent events.
Our proposal is the first to address this.

\textbf{Physical Layer Fingerprinting:} Radio frequency fingerprinting~\cite{dongre2021message} can fingerprints the transmitting (Tx) devices based on their physical layer characteristics, e.g., frequency offset stemmed from imperfect hardware.
This can be used to detect the messages sent from the same Tx devices even though they have different/changing higher layer network identifiers (e.g., IP addresses or MAC addresses), which can further efficiently filter out these excessive rate of bogus messages sent by adversarial Tx devices.
However, there is no evidence that fingerprinting can be done efficiently, which could be even computationally heavier than signature verification.
Moreover, fingerprinting approaches could be different for different communication technologies and obfuscation on the physical layer~\cite{givehchian2023practical} could counter solutions relying on physical layer fingerprinting.
Our proposal achieves resilience to \ac{DoS} by augmenting software implementations without reliance on lower layer technologies that require modification of \ac{OBU} communication modules and warrant a separate future investigation.

\textbf{Solutions for Other Domains:} Human effort or client computation resource can be involved to defend against \ac{DoS} attacks.
Captcha-based solutions are hard to machine automate as they require user input on their interfaces~\cite{von2008recaptcha}.
It is straightforward that requiring user action or input in frequent \ac{V2V} communication is unrealistic, especially when driving a car.
Puzzle-based schemes~\cite{sun2017privacy,liu2018mitigating} can protect against \ac{DoS} attacks on \ac{V2I} communication for pairwise connections, but do not fit in the context of safety beacons and event-driven messages, which require low latency in connectionless \ac{V2V} communication.
Solutions against physical layer jamming~\cite{twardokus2022vehicle} are orthogonal to our scheme and can co-exist with our scheme.

%% file: section_compsec/problem.tex
\section{System and Adversary Models}
\label{sec:problem}

In this section, we explain system and adversary model we consider in this paper, illustrated in~\cref{fig:system}.

\subsection{System Model}

Vehicles, termed \emph{nodes}, are equipped with \acp{OBU}.
Nodes exchange messages over multiple network interfaces, e.g., \ac{DSRC}~\cite{campvsc2,kenney2011dedicated} and possibly intermittently, the Internet through WiFi or cellular networks; their clocks are synchronized through \ac{GNSS} modules or \ac{NTP} servers.
Nodes are issued short-term credentials, \acfp{PC} by a \ac{VPKI}~\cite{papadimitratos2008secure,kargl2008secure,papadimitratos2009vehicular,khodaei2014towards,khodaei2015key,khodaei2018secmace}, and sign their messages with the corresponding private keys.
Upon a \ac{PC} change, all protocol stack identifiers, including IP and media access control addresses, change~\cite{petit2015pseudonym}.

Each vehicle broadcasts beacons at a rate, $\gamma$, with the maximum rate is $\gamma_{max} = 10 Hz$ (i.e., 10 beacons per second).
Event-driven messages can include not only standardized \acp{DENM}~\cite{etsidenm} but also any \ac{V2V} messages triggered by specific events (including misbehavior evidence that we introduce in Sec.~\ref{sec:scheme}).
Due to their event-driven nature, they do not follow specific time patterns, while an event message can be repeated more than once based on its criticality~\cite{etsidenm}.
All received (benign) messages must be verified to ensure \ac{VC} security.

\subsection{Adversary Model}

\textbf{External adversaries:} We are concerned with clogging \ac{DoS} attacks exploiting computationally expensive signature verification (or validation; the two terms are used interchangeably).
We do not dwell on \ac{DoS} attacks on the physical or medium access control layers.
We focus on single-hop transmissions; multi-hop \ac{VC} transmissions (e.g., Geocast packets~\cite{festag2010design}) are out of scope.
An adversary can flood with bogus messages carrying fake signatures (i.e., random bits with same lengths as authentic signatures that are very fast to generate), irrespective of whether the adversary is a legitimate node or not~\cite{dong2013providing}.
We do not consider internal adversaries flooding with properly authenticated messages, because an abnormally high message frequency can be trivially detected and attributed to each \ac{PC} and eventually the long-term sender identity, followed by eviction~\cite{khodaei2018secmace}.

An attacking node that floods bogus messages utilizing its full bandwidth can broadcast, for example, at a rate higher than 2000$Hz$ considering a typical 6Mbps bandwidth for IEEE 802.11p~\cite{teixeira2014vehicular,chang2015v2v} and 300 bytes \ac{V2V} messages~\cite{calandriello2011performance}.
Attackers attach (random) bogus signatures and plausible \acp{PC} so that receiving nodes have to deem the messages useful and verify the signatures.
In the same spirit, they set randomized IP and media access control addresses for each beacon they transmit, so that receiving nodes cannot easily impose rate control based on the addresses.
This is equivalent to an aggregate message rate from 200 neighboring vehicles broadcasting at $\gamma = 10$ messages per second, i.e., 10Hz.
Multiple such attacking nodes can be deployed (simple adversary controlled devices or malware on infected \acp{OBU}) in a targeted area, essentially launching a distributed \ac{DoS} attack.
All benign nodes within the communication range are victims and have to proceed with signature verification to check validity of the messages.
Unlike defenses against \ac{DoS} towards client-server based architectures, e.g.,~\cite{sun2017privacy,liu2018mitigating}, a \ac{DoS}-resilient \ac{V2V} communication scheme should not introduce significant overhead or delay on the sending process that would degrade the overall performance of \ac{VC} applications.

\textbf{Internal adversary:} Compromised nodes, termed \emph{malicious nodes} in the rest of the paper, are equipped with valid credentials, and can actively inject authenticated false data to the network, in order to mislead other nodes.
This is particularly relevant for cooperative verification: one adversary could falsely inform its neighbors that it validated bogus beacons transmitted in the same neighborhood.
We are concerned with message signature validity, while we do not dwell on actual location or content verification on the authenticated beacons; these can be addressed with position verification~\cite{FioreCCP:J:2013,PoturalskiPH:J:2013} or content validation~\cite{raya2008data,gisdakis2015shield} approaches.

\subsection{Requirements}

\textbf{Message and entity authentication:} The message sender must be (pseudonymously) authenticated, allowing receivers to corroborate the sender legitimacy and verify they were not altered or replayed.

\textbf{Non-repudiation and accountability:} Nodes should not be able to deny actions performed, thus messages sent.
They are accountable for their messages or actions, and any message transmission mandated our scheme should be non-repudiable.
Any misbehaving node should have their long-term identities revealed and, if needed, evicted from the system.

\textbf{Ano-/Pseudo-nymity and unlinkability:} Messages should not be linkable to their sender's long-term identity. They should be linkable only over a protocol-selectable period, i.e., over a pseudonymous identity validity period.

\textbf{Availability:} Nodes should maintain their ability to timely validate legitimate messages (\acp{CAM} and \acp{DENM}) even amidst an adversarial flood of fictitious traffic.
We do not impose strict requirements on message verification deadline.
Rather, the verification delay should be within the order of magnitude of message dissemination intervals (e.g., beacon intervals or event triggering intervals), in order to ensure persistent neighbor awareness and timely environment notification awareness, given certain extent of packet loss and vehicle mobility predictability.

%% file: section_compsec/scheme.tex
\section{Proposed Scheme}
\label{sec:scheme}

\subsection{Overview}

For timely validation of safety beacon and event-driven messages even under \ac{DoS} attacks, our scheme extends pseudonymous authentication with several DoS-resilient features: (i) efficient (bogus) beacon filtering based on hash chains, (ii) cooperative verification, (iii) self-chained verification, and (iv) event message verification facilitators.
We trade off communication overhead of such message verification facilitators for faster message verification.
To enhance security, our scheme detects misbehaving nodes by probabilistically checking and cross-checking messages, to render cooperative verification robust to internal adversaries.
\Cref{table:notation} summarizes the notation used in the rest of the paper.
	
	\begin{table}[t]
		\caption{Notation}
		\centering
		\resizebox{0.45\textwidth}{!}{
		\begin{tabular}{l | *{1}{c} r}
			\hline \hline
			$PC$ & \emph{\acl{PC}} \\\hline
			$PRL$ & \emph{\acl{PRL}} \\\hline
			$KRL$ & \emph{\acl{KRL}} \\\hline
			$\{msg\}_{\sigma_{PC}}$ & \emph{Signed message attached signature and \ac{PC}} \\\hline
			$Pr_{check}$ & \emph{Probability of checking cooperative verifier} \\\hline
			$\alpha$ & \emph{Number of message facilitators in a beacon} \\\hline
			$\beta_1$ & \emph{Event message facilitator repeat counter} \\\hline
			$\beta_2$ & \emph{Event message repeat counter} \\\hline
			$\gamma$ & \emph{Beacon frequency} \\\hline
			$k$ & \emph{Number of self-chained verifiers in a beacon} \\\hline
			$\tau$ & \emph{Average message verification delay} \\\hline
			$T_{blife}$ & \emph{Beacon lifetime} \\\hline
			$H()$ & \emph{Hash function} \\\hline
			$MAC\ /\ MAC_K(msg)$ & \emph{\acl{MAC} value / $H(K\ ||\ msg)$} \\\hline
			$Queue_{recv}$ & \emph{Beacon reception queue} \\\hline
			$Queue_{check}$ & \emph{Checked beacon queue} \\\hline
			$v\ /\ Ver$ & \emph{Beacon verifier / Beacon verifier set} \\\hline
			$f\ /\ F$ & \emph{Message facilitator / Message facilitator set} \\\hline
			\hline
		\end{tabular}
		\renewcommand{\arraystretch}{1}
		\label{table:notation}
		}
	\end{table}

\begin{figure}[t]
	\centering
	\includegraphics[width=0.9\linewidth]{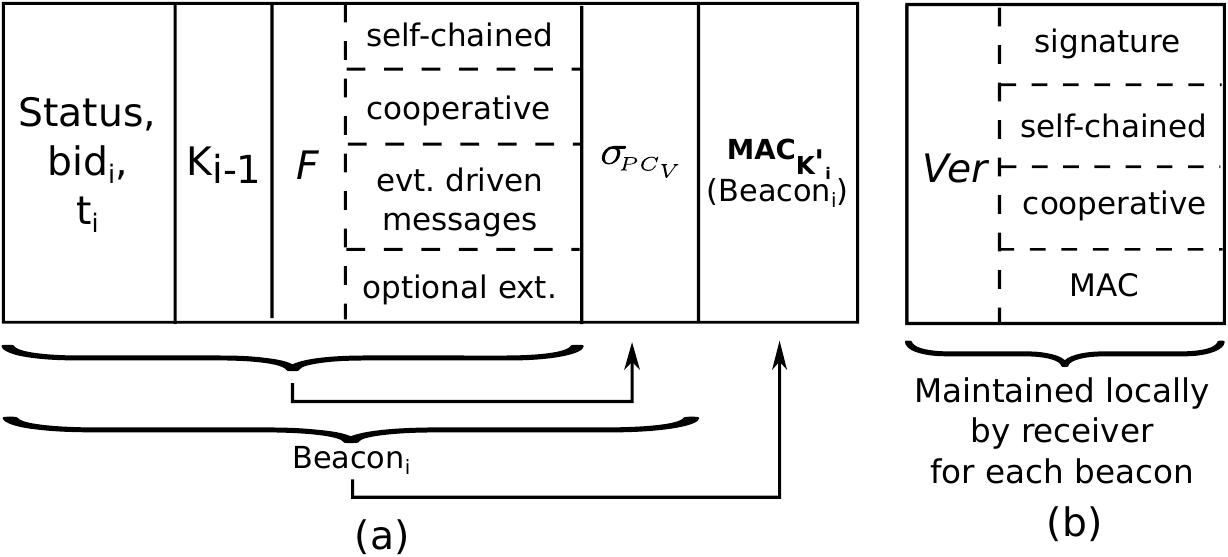}
	\caption{(a) The extended beacon format and (b) the locally maintained beacon verifier set.}
	\label{fig:format}
\end{figure}

\textbf{Tracing benign beacons (and filtering out bogus beacons):} Nodes in our scheme are issued short-term \acp{PC} by the \ac{VPKI}.
We consider a Sybil-resilient \ac{PC} lifetime policy~\cite{khodaei2014towards,khodaei2018secmace}, so that each node is equipped with \acp{PC} with non-overlapping lifetimes.
Nodes authenticate messages by signing with their private keys corresponding to the \acp{PC}. 
In addition, nodes maintain a key chain (essentially a hash chain) for each of their \acp{PC}.
Each key chain element is used as a (symmetric) \textit{one-time beacon authentication key} (\cref{subsec:chain}), authenticating strictly one beacon that broadcasted in the corresponding time interval.
These key chains, once verified, assist the receiving nodes in filtering out bogus beacons at a lower cost than performing signature verifications (\cref{subsec:reception}).
Sending nodes piggyback key chain elements to their disseminated beacons and the receivers queue or drop the beacons based on correctness of the one-time keys.
The extended beacon piggybacking one-time key, as shown in \cref{fig:format}, is:
\begin{align}
	B_i = \{ S, bid_i, t_i, K_{i-1}, F \}_{\sigma_{PC}}.
\end{align}
It comprises ``traditional'' beacon fields, e.g., vehicle status ($S$, including location, velocity, direction, etc.), beacon id ($bid_i$) and timestamp ($t_i$), the corresponding one-time key ($K_{i-1}$), and message facilitators ($F$, explained below).
Each signed beacon carries a one-time keyed \ac{MAC}.
Unlike TESLA~\cite{perrig2000efficient} based beacon authentication~\cite{studer2009flexible,lyu2016pba}, our scheme merely uses validation of their \acp{MAC} to filter out bogus beacons and cross-check the correctness of cooperative verification (\cref{subsec:mac}), while message acceptance in our scheme relies on signature verification, self-chained verification or cooperative verification.
The actual broadcasted beacon message is:
\begin{align}
M_i = \{ B_i, MAC_i \}.
\end{align}

\begin{figure*}[t]
	\centering
	\includegraphics[width=0.65\linewidth]{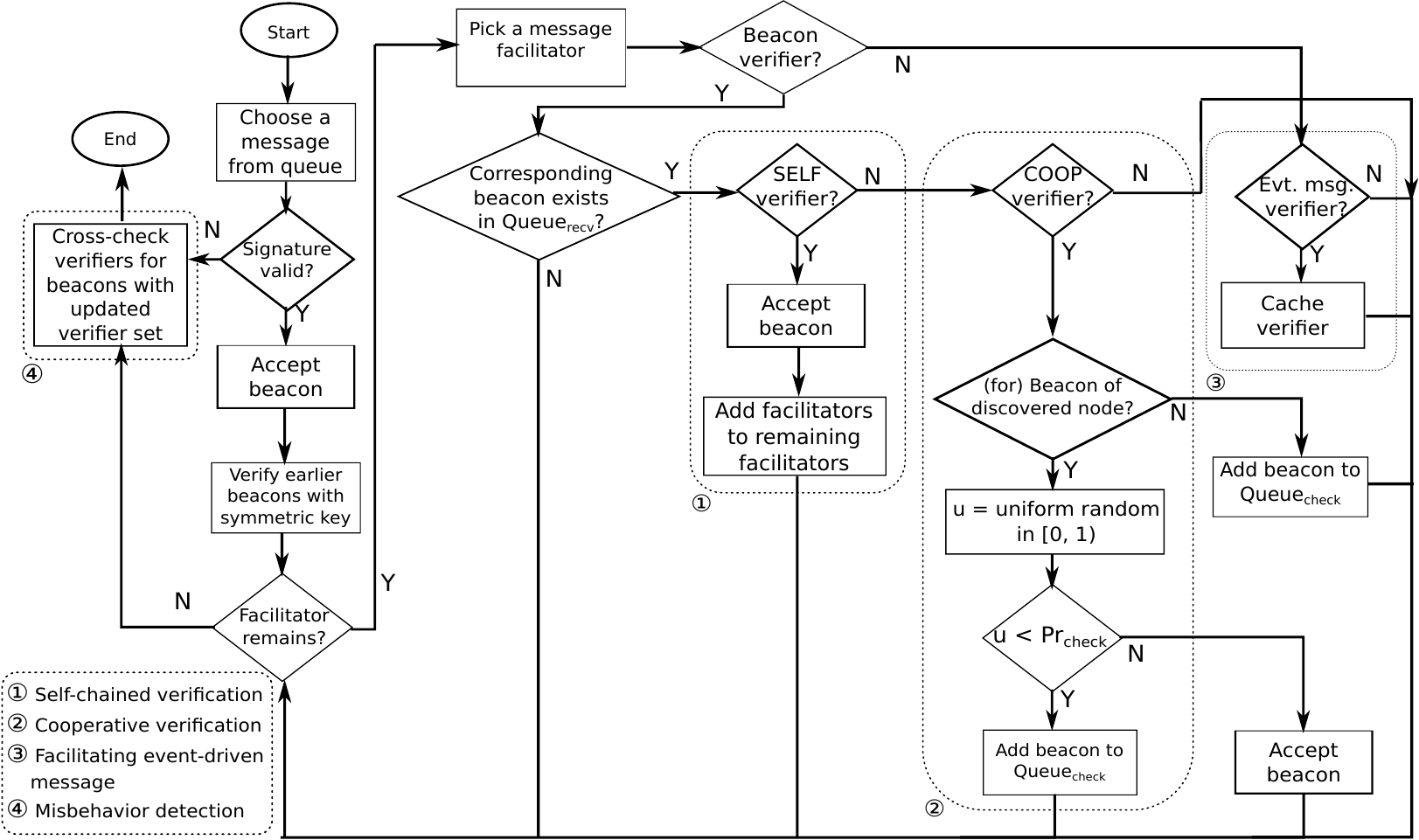}
	\caption{Flowchart of message verification.}
	\label{fig:diagram}
\end{figure*}
	
\textbf{Expediting node discovery and beacon verification:} Tracing and filtering beacons, however, is possible only after key chains are verified, i.e., at least one signed beacon (that carries a one-time key) from a neighboring node is verified (thus the node is discovered).
To expedite node discovery (and, in general, beacon verification), apart from continuous queued beacon verification (\cref{subsec:verification}), nodes share \textit{cooperative verifiers} (as one type of message facilitator) that point to beacons they verified based on signatures.
Sharing cooperative verifiers help neighbors to validate beacons or discover new neighbor nodes.
More specifically, received cooperative verifiers could point to locally queued beacons of either non-discovered nodes or discovered nodes.
The former ones are kept in a special-purpose queue, used for node discovery: the beacons in this queue wait to be verified based on signatures.
The latter ones validate corresponding beacons with probabilistic signature checking, in order to detect any internal adversary that disseminate fraudulent cooperative verifiers for previously broadcasted bogus beacons.
Strictly performing signature verification for non-discovered nodes' beacons is important, because this sets a basis for subsequent efficient beacon verifications, establishing trust on both the vehicle status and key chain.

Apart from the signature verifier (the signature itself) and the cooperative verifier, beacon verification can also rely on the self-chained verifier.
Each beacon carries the latter, pointing to messages disseminated in the immediate past, to verify more than one beacons by the same sender upon a single signature verification.

\textbf{Misbehavior detection:} The redundant verifiers for the same message, maintained by the receiver in a local data structure, $Ver$ (\cref{fig:format}), are cross-checked to detect misbehavior.
The MAC, while not sufficient for non-repudiable beacon verification, can be used as an extra reference to cross-check any former verifier.
If the verifiers conflict, the beacon signature must be examined, to reveal the faulty verifiers.
Signature or self-chained verifiers both suffice to establish message validity, because they both rely on the verification of signature generated by the message sender.
The source of a faulty verifier is blacklisted locally (and reported to the \ac{VPKI}).

\textbf{Facilitating event-driven messages:} Beacons piggyback event message facilitators for \ac{DoS}-resilient event message verification (\cref{subsec:event}).
A facilitator is essentially the hash value (thus verifier) of an event message, but it is not used to verify the message.
Rather, each event message facilitator is cached locally to wait for matching (artificially delayed) event message.
An event message that does not match any locally cached facilitator is dropped.
Event messages are always verified based on signatures, to ensure non-repudiation of the claimed events.
Such matching based on proactive facilitator dissemination can help receivers to efficiently catch potentially valid event-driven messages, while filtering out irrelevant bogus messages.

We show a high-level flowchart of our scheme with \cref{fig:diagram}, before detailed explanation on each scheme component.
Each node chooses a beacon to verify from the local queue, $Queue_{recv}$ and $Queue_{check}$, and verifies the signature.
We explain the purpose of the two queues when we present in detail the beacon verification below (\cref{subsec:component}).
If the signature is valid, MACs of previous beacons (if any not verified) by the same sender are verified based on the symmetric key in this beacon.
Each facilitator can be used to verify self-chained beacons or cooperatively validated beacons, or simply stored to wait for matching event-driven messages.
For each cooperatively validated beacon, if the beacon sender was not discovered, then this first-validated beacon serves for node/neighbor discovery; otherwise, the beacon is accepted or probabilistically checked based on a probability $Pr_{check}$.

\subsection{Scheme Components}
\label{subsec:component}

\subsubsection{Beacon Chaining}
\label{subsec:chain}

\begin{figure}[t]
	\centering
	\includegraphics[width=0.9\columnwidth]{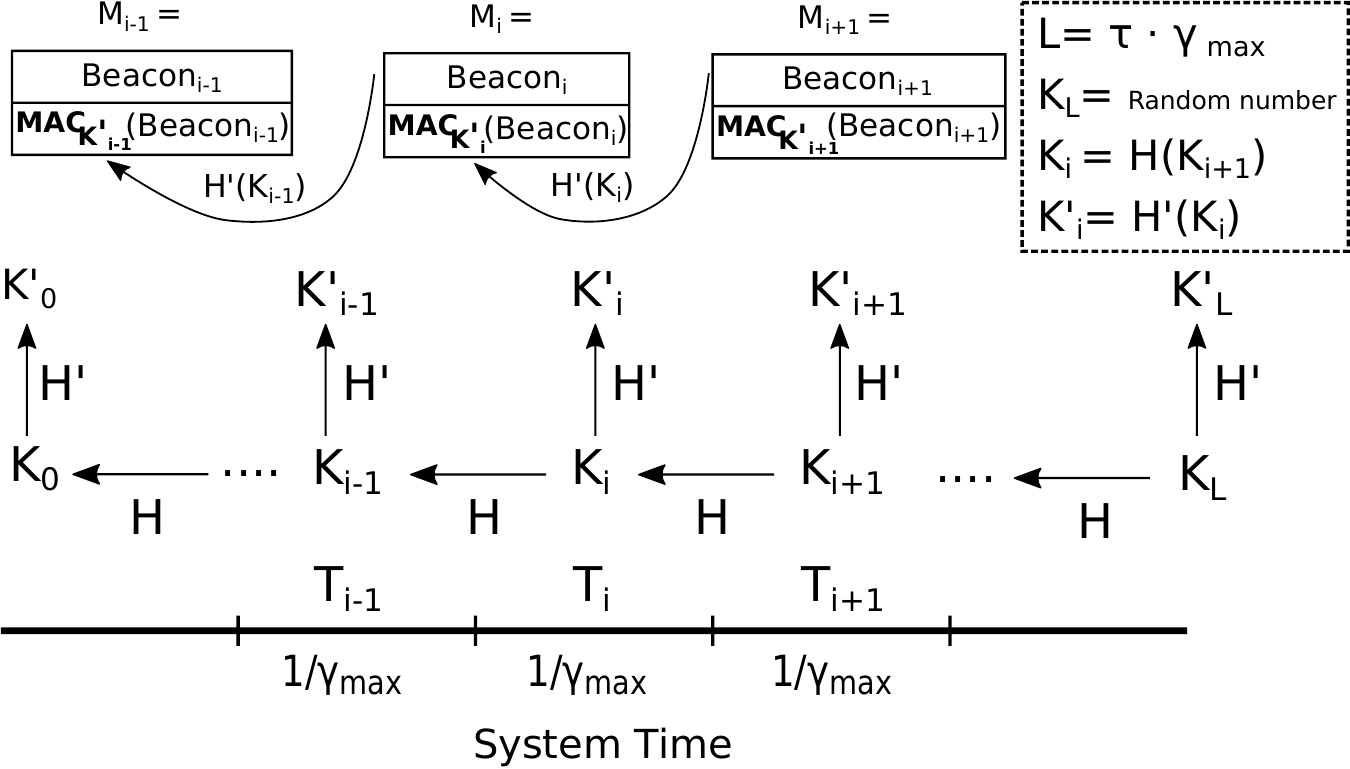}
	\caption{Beacon chaining.}
	\label{fig:mac}
\end{figure}

A unique key chain is generated by each node for each own \ac{PC}.
A key chain pertains to a \ac{PC} lifetime and broadcasted beacons throughout the \ac{PC} lifetime.
The length of a key chain is the number of beacon that can be broadcasted under a \ac{PC}.
With $\gamma_{max}= 10 Hz$, the length of a key chain $L= \tau \cdot \gamma_{max}$.
Therefore, it guarantees that the key chain elements are sufficient for authenticating all beacons that could be disseminated with the corresponding \ac{PC}.
\Cref{fig:mac} shows key chain generation and beacon chaining process.
The system time is divided into equally-sized time slots, and the length of each time slot is $1/\gamma_{max}$ (i.e., 100 $msec$): the shortest possible beacon interval according to the standard~\cite{cam}.
Each $K'_{i}$ is used as an authentication key to compute a \ac{MAC} for a beacon disseminated within the time slot, $T_{i}$, and the corresponding $K_{i}$ is disclosed with the next beacon in the next time slot, $T_{i+1}$.
Therefore, only the key chain owner has access to the key that can be used to authenticate the beacon in the current time interval/slot.
Once a symmetric key is disclosed, it can be only used to verify \acp{MAC} of earlier beacons, while any message authenticated with a disclosed symmetric key (by attackers) will be dropped.
The trust on key chain is established based on signature verification.
Once a beacon is verified, the one-time key on that beacon can be trusted and the rest of key chain elements can be verified.

In general, key chain generation is not an issue, considering the low computation overhead of hash functions. Consider a vehicle pre-generates key chains for a day, while parked during the midnight. The vehicle requires at most $864,000$ one-time keys for a day (i.e., $24$ $h$, while an actual trip duration during the day could be much shorter) with $\gamma = \gamma_{max}$. Thus, the maximum required storage capacity is $864,000 * Size_{digest}$. With a $Size_{digest} = 160bits$, (e.g., SHA-1\footnote{Although collision attack has been successfully applied on SHA-1~\cite{stevens2017first}, our scheme relies on second-preimage resistance of hash function, which is much harder to attack.}), the required storage would be only around $14$ $MB$, not an issue for storage in modern mobile device.

\subsubsection{DoS-resilient Beacon Reception}
\label{subsec:reception}

\begin{algorithm}[t]
	\caption{DoS-resilient Beacon Reception}
	\label{alg:recv}
	\textbf{Input:} $\{B_i, MAC_i \} = M_i$ \\

	$\{ S, bid_i, t_i, K_{i-1}, F \}_{\sigma_{PC}} = B_i$ \\
	Check $PC$ against PRL; proceed or drop $M_i$ accordingly. \\
	$M_i = \{B_i, MAC_i, Ver_{B_i} = \{\}\}$ \\
	\eIf {$PC \in PC\_Discovered$}{
		$\{PC, t_{i'}, K_{i'-1}, Bs \} = Search(PC, PC\_Bs)$\\
		\eIf {$t_i \in T_i$ \textbf{and} $H_{C}^{i-i'}(K_{i-1}) == K_{i'-1}$}{
			$Queue_{recv} = \{M_i\} \cup Queue_{recv}$ \\
			Update $PC\_Bs$ with $\{PC, t_{i}, K_{i-1}, \{M_i\} \cup Bs\}$. \\
			$M^\prime = $ the latest beacon from $Bs$\\
			\If {$M^\prime \neq \{\}$}{
				Input $M^\prime$ to \cref{alg:mac} \\
			}
		}{
			Drop $M_i$.\\
		}
	}{
		\eIf {$PC \in PC\_Bs$}{
			$\{PC, nil, nil, Bs \} = Search(PC, PC\_Bs)$ \\
			\eIf {$K_{i-1} == $ any one-time key in $Bs$}{
				Drop $M_i$ \\
			}{
				$Queue_{recv} = \{M_i\} \cup Queue_{recv}$ \\
				Update $PC\_Bs$ with $\{PC, nil, nil, \{M_i\} \cup Bs\}$. \\
			}
		}{
			$Queue_{recv} = \{M_i\} \cup Queue_{recv}$ \\
			$PC\_Bs = PC\_Bs \cup \{\{PC, nil, nil, \{M_i\}\}\}$
		}
	}
\end{algorithm}

Nodes handle received beacons according to \cref{alg:recv}, before their actual signature or cooperative verifications.
Each node maintains a local \ac{PRL} that includes both periodically downloaded \ac{PRL} from the \ac{VPKI} and \acp{PC} of locally detected misbehaving nodes (not yet announced with the latest \ac{VPKI} \ac{PRL}).
Trust on a key chain can be established once a beacon is verified based on its signature, and we consider the node as discovered.
The DoS-resilient node discovery process is illustrated with \cref{alg:coop} (\cref{subsec:coop}).

Once a beacon is received, the node checks whether the attached \ac{PC} is revoked or not (\ref{alg:recv}:1-3, i.e., lines 1-3 in \cref{alg:recv}).
Beacon messages that pass the \ac{PRL} checking are extended with an empty verifier set (\cref{fig:format}), $Ver_{B_i}$ (\ref{alg:recv}:4), storing upcoming verifiers to be cross-checked.
There are four types of verifiers: signature verification (termed $SIG$ verifier), self-chained ($SELF$) verifier, cooperative ($COOP$) verifier, and $MAC$ verifier; we explain each verifier with in \cref{subsec:verification,subsec:mac,subsec:self,subsec:coop}.
If the \ac{PC} owner was already discovered, the receiver fetches the latest cached information (i.e., $\{PC, K_{i'-1}, t_{i'}, Bs \}$) for the \ac{PC} (\ref{alg:recv}:5-6).
$PC\_Bs$ stores information corresponding to each discovered or non-discovered nodes.
For each discovered node, it stores the latest one-time key ($K_{i'-1}$) and timestamp ($t_{i'}$) fetched from the latest beacon, and the list of non-expired beacons ($Bs$) that successfully ``passed'' hash chain checks (explained below).
For the non-discovered nodes, it stores only the beacon list, and the other two fields are updated once the node is discovered (if any benign beacon exist in the beacon list).
With the received beacon, the receiver checks the correctness of $K_{i-1}$ against $K_{i'-1}$: the number of intervals between the two keys should be consistent with the difference between $t_i$ and $t_{i'}$ (\ref{alg:recv}:7).
This preliminary hash chain check can efficiently drop bogus beacons attached incorrect or replayed overheard disclosed one-time keys.
The beacon is then queued for verification and the information in $PC\_Bs$ is updated (\ref{alg:recv}:8-9).
The \ac{MAC} of the latest beacon (before this reception) in $Bs$ is checked with \cref{alg:mac} (\ref{alg:recv}:10-12; see \cref{subsec:mac}).
Any received beacon, that fails hash chain test, is simply dropped (\ref{alg:recv}:13-14).
If the sender node is not yet discovered, the receiver checks whether the piggybacked one-time key was seen from any previous beacon attached the same \ac{PC}; if not, the beacon is added to $Bs$ of the \ac{PC} and queued for verification (\ref{alg:recv}:16-25).

Although actual beacon lifetime (the time duration each beacon stays in \ac{OBU} upon reception) depend on various system parameters (e.g., content relevance and beacon rate), without loss of generality, we assume each beacon is given a lifetime, $T_{blife}$.
Without such a lifetime, \ac{OBU} memory may be saturated by outdated beacons, which could never be verified due to higher message arrival rate than message processing rate.
Both newly received beacons, and verified beacons are kept for $T_{blife}$ to cross-check verifiers for the same beacon.
Moreover, existing schemes, e.g., for content verification~\cite{raya2007eviction} and adaptive beacon rate~\cite{nguyen2017mobility}, rely on message redundancy, thus require the verified messages to be stored for an extra period.

\subsubsection{Beacon Verification}
\label{subsec:verification}

\begin{algorithm}[h!]
	\caption{Beacon Verification}
	\label{alg:vrfc}
		\While {$M = \{\}$ \textbf{and} $Queue_{check}$ is not empty}{
			$\{B_{i}, MAC_{i}, Ver_{B_i}\}$ = $Queue_{check}$ head, $M$ \\
			\If {$M$ is only for node discovery \textbf{and} $PC$ of $M$ $\in$ $PC\_Discovered$}{
				\If{$Ver_{B_i}$ contains positive COOP verifier }{
					Accept $M$ with prob. $1-Pr_{check}$;  $M = \{\}$.
				}
			}
		}
		\If {$M \neq \{\}$ \textbf{and} $Queue_{recv}$ is not empty}{
			$Condition: t_{beacon} + 1/\gamma_{max} > t_{now}$ \\
			$Beacon\_Set_{D/ND} = $ discovered/non-discovered nodes' beacons in $Queue_{recv}$ that meet $Condition$ \\
			\uIf {$Beacon\_Set_{D} \cup Beacon\_Set_{ND}  == \phi$} {
				$M$ =  $Queue_{recv}$ head \\
			}
			\uElseIf {$time_{D} / (time_{D} + time_{ND})  <= Ratio_{D}$ \textbf{or} $Beacon\_Set_{ND}  == \phi$} {
				$M = $ randomly chosen from $Beacon\_Set_{D}$ \\
			}
			\Else{
				$M = $ randomly chosen from $Beacon\_Set_{ND}$ \\
			}
		}
		\If {$M \neq \{\}$} {
				$\{B_i, MAC_i, Ver_{B_i}\} = M$\\
				$\{S, t_i, K_{i-1}, F\}_{\sigma_{PC}} = B_i$ \\
				$validity$ = $B_i$ signature validity; update $time_{D/ND}$. \\

			\If {$validity = true$} {
				Accept $M$. \\
				\If {$PC \notin PC\_Discovered$}{
					Add PC to $PC\_Discovered$. \\
					$\{PC, nil, nil, Bs \} = Search(PC, PC\_Bs)$ \\
					\For {each $M_j$ in $Bs$ from head}{
						\eIf{$M_j$ is $Bs$ head}{
							$\{B_{j}, MAC_{j}, Ver_{B_j}\}$ = $M_j$ \\
							$\{ S, bid_{j}, t_{j}, K_{j-1}, F \}_{\sigma_{PC}}$ = $B_j$ \\
							\eIf {$t_i$ is within $T_i$ period \textbf{and} $H_{C}^{i-j}(K_{i-1}) == K_{j-1}$}{
								Update $PC\_Bs$ with $\{PC, t_{j}, K_{j-1}, Bs\}$. \\
							}{
								Drop $M_j$.\\
							}
						}{
							Input $M_j$ into \cref{alg:mac}. \\
						}
					}
				}
				Input $\{M_i, SIG\}$ into \cref{alg:self,alg:coop}. \\
			}
			$v_{new} = \{nil, nil, validity, SIG\}$ \\
			$Ver_{B_i} = \{v_{new}\}$ \\
			
		}
\end{algorithm}

Each node maintains two queues: $Queue_{recv}$ and $Queue_{check}$. $Queue_{recv}$ queues newly received beacons for a \emph{semi-randomized} (explained below) \ac{LCFS} verification, and $Queue_{check}$ stores potentially valid beacons from non-discovered nodes and beacons chosen for probabilistic checking.
$Queue_{check}$ is given higher priority for quicker node discovery and misbehavior detection. We choose the \ac{LCFS} strategy for beacon verification, because our scheme can efficiently validate older beacons based on self-chained verifier once a newer beacon is verified, and fresher cooperative verifiers can benefit neighboring nodes to a greater extent, i.e., verifying latest received beacons.
Moreover, with high bogus beacon rate under \ac{DoS} attacks and a given beacon lifetime, \ac{LCFS} verification guarantees that fresh beacons are verified, instead of the oldest non-expired beacons with a \ac{FCFS} verification.

If $Queue_{check}$ is not empty, the beacon at $Queue_{check}$ head is chosen for verification.
If the beacon was chosen for node discovery, but the beacon sender was already discovered while waiting in $Queue_{check}$, then the beacon is simply accepted with a probability $1-Pr_{check}$; otherwise, if not yet discovered, the protocol proceeds with signature verification (\ref{alg:vrfc}:3-5).
If the first phase (\ref{alg:vrfc}:1-5) did not conclude with a beacon for verification, then a beacon will be chosen from $Queue_{recv}$.

In order to guarantee timely benign beacon verification under \ac{DoS} attacks, we assign explicitly CPU time ratios for verifying discovered nodes' and non-discovered nodes' beacons. 
Discovered nodes' beacon verifications occupy a time ratio of $Ratio_{D}$, and non-discovered nodes' beacon verifications occupy a time ratio of $Ratio_{ND}$, where $Ratio_{D} + Ratio_{ND} = 1$).
$time_{D}$ and $time_{ND}$ maintain time used for verifying beacons from each category. Moreover, a beacon is randomly chosen among beacons that fulfill the condition: $t_{beacon} + 1/\gamma_{max} > t_{now}$, where $t_{beacon}$ is the original timestamp on each beacon.
This time condition ensures that cooperative verifiers of relatively fresher beacons are piggybacked; the randomization maximizes the benefit from cooperative verifiers, reducing the chance closely located vehicles verifying (thus piggybacking the cooperative verifiers to) the same beacons, which would be the case if beacons are strictly chosen from their $Queue_{recv}$ heads. The nodes choose a beacon from $Queue_{recv}$ based on the above design choices (\ref{alg:vrfc}:6-14).

If a beacon is chosen from $Queue_{check}$ or $Queue_{recv}$, the signature is verified (\ref{alg:vrfc}:15-18).
If the valid beacon belongs to a non-discovered node, \ac{PC} is added to $PC\_Discovered$ (\ref{alg:vrfc}:19-22), and all queued beacons attached $PC$ are then checked with MAC (\ref{alg:vrfc}:23-24,33).
The latest beacon attached the correct one-time key is only used to update the cached information (\ref{alg:vrfc}:25-29), because a newer beacon is necessary for its MAC validation.
The valid beacon is then used for self-chained verification and cooperative verification (\ref{alg:vrfc}:34), and the signature verification result is cross-checked with previous verifiers to detect any misbehaviors (\ref{alg:vrfc}:35-36).
After the conflict checking, this signature verification result replaces all previously stored verifiers, because the new verifier is the definitive $SIG$ verifier and any other verifier is unnecessary.

\textbf{Cooperative verifier maintenance:} Each node maintains a list of recently verified beacons based on signatures, to be piggybacked on own beacons as cooperative verifiers. The $\alpha$ latest (in terms of their timestamps) verified beacons are kept.
Each of the $\alpha$ beacons could be either a valid beacon, or a bogus beacon that attached discovered node's \ac{PC} and correct one-time key, i.e., a cooperative verifier could be either positive or negative.
A negative cooperative verifier can potentially reveal a misbehavior that malicious nodes attempt to validate the corresponding bogus beacon with a false positive cooperative verifier, because the two cooperative verifiers would conflict.
We provide qualitative and quantitative security analysis for the effect of negative cooperative verifiers on malicious node detection in~\cref{sec:analysis,sec:simulation}.

\subsubsection{MAC Validation}
\label{subsec:mac}

\begin{algorithm}[t]
	\caption{MAC Validation}
	\label{alg:mac}
	
	\textbf{Input:} $\{B_i, MAC_{i}, Ver_{B_i} \} = M$ \\
	$\{S, t_i, K_{i-1}, F\}_{\sigma_{PC}} = B_i$ \\
	$\{PC, t_{i'}, K_{i'-1}, Bs \} = Search(PC, PC\_Bs)$\\
	$validity$ =  ($H_{C}^{i-i'}(K_{i-1}) == K_{i'-1}$ \textbf{and} $MAC_{H'(K_{i'-1})}(B_i) == MAC_i$) \\
	\If {$validity = true$ \textbf{and} $M$ was not validated \textbf{and} $PC \notin KRL$}{
		Input $\{M, MAC\}$ to \cref{alg:coop}. \\
		$v_{new} = \{pcid_{PC}, bid_{i'}, true, MAC\}$ \\
		$Ver_{B_i} = Ver_{B_i} \cup \{v_{new}\}$ \\
	}
\end{algorithm}

MACs of discovered nodes' beacons are checked once newer beacons that piggybacked correct one-time keys are received (\cref{alg:mac}). Each node maintains a \ac{KRL} that stores all detected node \acp{PC} attempting to misuse one-time key beacon authentication by attaching correct one-time keys and MACs but bogus signatures.
In this algorithm, the MAC correctness is checked first, and, if validated and \ac{PC} is not in \ac{KRL}, the $COOP$ verifiers in the beacon is used to find potentially valid beacons for non-discovered nodes (\ref{alg:mac}:4-6). The beacon verifiers are cross-checked if the beacon is not already in $Queue_{check}$, and the new $MAC$ verifier is added to the verifier set (\ref{alg:mac}:7-8). If a \ac{PC} is in \ac{KRL}, even if the MAC checks are passed, the facilitators in the corresponding beacons will not be used. 

\subsubsection{Self-chained Verification}
\label{subsec:self}

\begin{algorithm}[h]
	\caption{Self-chained Verification}
	\label{alg:self}
	\textbf{Input:} $\{B_i, MAC_{i}, Ver_{B_i} \} = M$ \\
	$\{S, t_i, K_{i-1}, F\}_{\sigma_{PC}} = B_i$ \\
	$\{PC, K_{i'}, t_{i'+1}, Bs \} = Search(PC, Cache_{PC})$ \\
	\For {each $f$ in $F$} {
		\If {$f$ is $SELF$ verifier} {
			$\{bid, Digest\} = f$ \\
			$M' = Search(bid, Bs)$ \\
			\If {$M \neq \{\}$ \textbf{and} $M$ was not validated based on signature or SELF verifier } {
				$\{B_j, MAC_{j}, Ver_{B_j} \} = M'$ \\
				$validity = (H(B_j) == Digest)$ \\
				\If {$validity == true$} {
					Accept $M'$ \textbf{if} was not accepted. \\
					Input $\{M', SIG\}$ into \cref{alg:self,alg:coop}. \\
				}
				$v_{new} = \{pcid_{PC}, bid_i, validity, SELF\}$ \\
				$Ver_{B_j} = \{v_{new}\}$ \\
			}
		}
	}
\end{algorithm}

Each beacon is piggybacked with $k$ $SELF$ verifiers, pointing to immediate previous $k$ own beacons. A $SELF$ verifier is a tuple of beacon id ($bid$) and the corresponding beacon digest. For each $SELF$ verifier, the node checks whether a beacon with the same $bid$ is received (\ref{alg:self}:4-7). If such a beacon exists and was not verified based on $SIG$ or $SELF$ verifiers, the beacon digest is compared with the one in the verifier (\ref{alg:self}:8-9). If they are equal, $M$ is accepted and used further for self-chained verification and cooperative verification (\ref{alg:self}:10-12). The $SELF$ verifier is used to cross-check existing verifiers for misbehavior detection (\ref{alg:self}:13-15). Only this $SELF$ verifier is kept (same as $SIG$ verifier in \cref{alg:vrfc}), because it is equivalent to signature verification: the beacon signature corroborates the current beacon and the previous $k$ own beacons.

\subsubsection{Cooperative Verification}
\label{subsec:coop}

\begin{algorithm}[t]
	\caption{Cooperative Verification}
	\label{alg:coop}
		\textbf{Input:} $\{B_i, type\}$\\
		$\{ S, bid_i, t_i, K_{i-1}, F \}_{\sigma_{PC}} = B_i$ \\
		\For {each verifier $f$ in $F$} {
			$\{pcid, bid, validity, Digest\} = f$ \\
			$\{PC, K_{i'}, t_{i'+1}, Bs \} = Search(PC, Cache_{PC})$ \\
			$M = Search(bid, Bs)$ \\
			\If {$M \neq \{\}$} {
				$\{B_j, MAC_{j}, Ver_{B_j} \} = M$ \\
				\If {$H(B_j) == Digest$} {
				\If{$M \in Queue_{recv}$}{
					Remove $M$ from $Queue_{recv}$. \\
					\uIf {$PC'\ (on\ $M$) \notin PC\_Discovered$ \textbf{and} $validity == true$} {
						$Queue_{check} = Queue_{check} \cup \{M\}$. \\
					}
					\ElseIf {$type == SELF\ or \ SIG$} {
						$Queue_{check} = Queue_{check} \cup \{M\}$ with a prob. of $Pr_{check}$; otherwise, accept/reject $M$ based on $validity == true/false$. \\
					}
				}
				$v_{new} = \{pcid_{PC}, bid_i, validity, COOP\}$ \\
				$Ver_{B_j} =  Ver_{B_j} \cup \{v_{new}\}$ \\
			}
			}
		}
\end{algorithm}

The $COOP$ verifier, points to beacons previously the beacon sender verified and can be used to verify matching beacons in $Queue_{recv}$ (\cref{alg:coop}).
Similar to $SELF$ verifier, each $COOP$ verifier is a tuple of \ac{PC} id ($pcid$), beacon id ($bid$), the claimed validity ($validity$), and the corresponding beacon digest (\ref{alg:coop}:3-6).
When a beacon, $M$, with matching $pcid$ and $bid$ is found, the hash value of $M$ is compared to the verifier (\ref{alg:coop}:7-9).
If $M$ is still in $Queue_{recv}$, the beacon is used for node discovery or cooperative verification and it is then removed from $Queue_{recv}$ (\ref{alg:coop}:10-11).
If the verifier is positive and the beacon is from non-discovered node, the beacon is added to $Queue_{check}$ for node discovery (\ref{alg:coop}:12-13).
If $M$ carried an already discovered node \ac{PC} and the current $COOP$ verifier was from a beacon verified based on $SIG$ or $SELF$ verifier, $M$ is further checked with a probability of $Pr_{check}$.
Otherwise, with probability $1-Pr_{check}$, $M$ is accepted or rejected based on the claimed $validity$ (\ref{alg:coop}:14-15).
Finally, the new verifier is cross-checked with the existing verifiers, if $M$ is not already in $Queue_{check}$ (\ref{alg:coop}:16-17).

\subsubsection{Misbehavior Detection}
\label{subsec:detect}

Misbehavior detection is achieved through cross-checking verifiers to the same beacon.
Whenever a new verifier of a beacon is added from any verification, it is compared against the existing ones.
If there is any conflict and the beacon hasn't been verified based on signature verification or self-chained verification, the beacon will be pushed to $Queue_{check}$ for signature verification.
Both signature verification and self-chained verification results can be used as the proof for beacon (in)validity.
Any proven bogus beacon verification attempt based on $COOP$ verifier will be reported to the authority, and the misbehavior be added to local \ac{PRL}.
However, $MAC$ verification is insufficient as a misbehavior evidence, because symmetric key based authentication does not provide non-repudiation (see \cref{fig:bogus}).
The misbehavior misusing $MAC$ will be added to \ac{KRL}, and the follow-up $MAC$s from the misbehavior will not be trusted.

\subsubsection{Event-driven Message Dissemination}
\label{subsec:event}

Our scheme can be readily used to facilitate event-driven message verification.
Due to the event-driven natural, nodes deem event messages highly critical and strictly verify message signatures.
Before dissemination of each actual event message, a facilitator for that message is disseminated with an immediate next beacon.
Here, the facilitator is an identifier to the event message, including the message digest. Once the beacon is validated based on any verifier, the event message facilitator can be cached. 
Next, when the actual event message is received, if the message hash value matches the cached message digest, the message is kept for signature verification.

Each event message facilitator is disseminated $\beta_1$ times with $\beta_1$ consecutive beacons.
After each beacon dissemination that carries the facilitator, actual event message is disseminated after the corresponding one-time key is disclosed, i.e., after the next beacon dissemination.
For each of $\beta_1$ repetitions, actual event messages are disseminated $\beta_2$ times to ensure successful delivery.
More specifically, event message facilitator is repeated $\beta_1$ times and actual event messages are repeated $\beta_1 * \beta_2$ times.
Both parameters are flexible and can be adjusted based on message criticality, network condition, etc.

We consider standardized \ac{DENM}~\cite{etsidenm} and misbehavior evidence as two examples of event-driven messages in our evaluation (see \Cref{sec:simulation}). Once a misbehaving \ac{PC} is detected, apart from reporting the evidence to the authority, it can also be disseminated to the neighboring nodes for quicker malicious node eviction.

%% file: section_compsec/analysis.tex
\section{Security and Privacy Analysis}
\label{sec:analysis}

We provide a qualitative security and privacy analysis of our scheme, before a simulation-based quantitative result in~\cref{sec:simulation}.

\subsection{Privacy}

Our scheme does not introduce any additional privacy concern compared to the standard~\cite{etsicam,etsidenm,calandriello2011performance}, in terms of message or (pseudonymous) identity linkability.
All messages are properly pseudonymously authenticated.
Messages, piggybacking one-time keys from the same key chain, are linkable, but they can already be trivially linked based on the attached same \ac{PC}.
Beacon chaining does not exceed the \ac{PC} lifetime/usage period.
Key chain lengths are aligned with \ac{PC} lifetimes, thus messages are still only linkable over a \ac{PC} lifetime.
Piggybacked cooperative verifiers imply correlations of nodes in terms of their geographical locations (i.e., nodes are within each other's communication range).
This is already explicitly available based on the location information included in their messages.
Event messages are linkable to their facilitator carriers, however, this was also possible based on the attached same \ac{PC}.

\subsection{Corroborating Legitimate Participation}

Vehicular credential verification is a fundamental component for cooperative awareness and safety in \ac{VC}.
All \acp{PC} are authenticated and issued by the \ac{VPKI}.
\acp{PC} cannot be linked to long-term vehicle identifiers, but it is important to prevent adversaries from introducing any phantom vehicles/nodes with bogus beacons.
If messages and the attached \acp{PC} are verified, the legitimacy of the neighboring vehicles can be proven.
Moreover, issuance of \acp{PC} with non-overlapping lifetimes ensure each vehicle is equipped at most one valid \ac{PC} at any point in time, thus preventing Sybil-based behavior.
In our scheme, beacons could be also accepted based on self-chained verifiers or cooperative verifiers.
Self-chained verifiers always correspond to already discovered nodes, so that they cannot be used to validate any beacon-attached non-verified \ac{PC}.
We prove below that adversaries are not able to introduce any phantom node by abusing cooperative verifiers, even though cooperative verifiers could point to third nodes.

\begin{theorem}
	\label{theorem:1}
	If the underlying public-key cryptography is secure and the \ac{VPKI} policy mandates issuance of \acp{PC} with non-overlapping lifetimes, an adversary cannot introduce phantom nodes/vehicles.
\end{theorem}

\begin{proof}
	In order to introduce a phantom node, an adversary has to disseminate a bogus beacon carrying a forged \ac{PC}.
	If such a beacon is chosen from $Queue_{recv}$ for signature verification, it will be rejected immediately due to the verification failure of the beacon signature.
	The adversary can attempt having the bogus beacon validated with a cooperative verifier piggybacked on an authentic beacon.
	The cooperative verifier would need to point to the bogus beacon.
	A benign node would insert the bogus beacon to $Queue_{check}$ based on the cooperative verifier.
	However, the bogus beacon will be rejected once the beacon signature verification fails.
	Therefore, any bogus beacon carrying a non-verified \ac{PC} will be proven invalid upon the signature verification and will be dropped (or it will simply expire). As a result, none of the receiving nodes will perceive this (fictitious \ac{PC} and sender).
\end{proof}

\subsection{Message Integrity and Authentication}

The standard-compliant pseudonymously authenticated \ac{V2V} communication entails message signing and verification with public/private key pairs that authenticated by the \ac{VPKI}, thus, provides message integrity and authentication.
Our scheme, while extending the beacon with additional fields, requires strict pseudonymous authentication on the messages, thus inheriting the message integrity and authentication.
Message timestamps prevent message replays, providing entity authentication upon successful message verifications.

\subsection{Non-repudiation and Accountability}

	Message content validation~\cite{raya2008data,gisdakis2015shield} is out of the scope here.
	However, upon detection of adversarial messages violating our scheme specification, internal adversaries should be held accountable.
	We are especially concerned with internal adversaries attempt to validate the bogus messages (attached bogus signatures) using message verification facilitators.

Given \Cref{theorem:1}, we know malicious nodes can only attempt validating bogus beacons, each attached a pair of valid \ac{PC} and correct one-time symmetric key, exploiting malicious self-chained verifiers or cooperative verifiers.
In order to take advantage of malicious cooperative verifiers, two or more malicious nodes need to collude.
A receiving node, once verified a properly signed beacon piggybacking malicious cooperative verifiers, might accept the corresponding bogus beacons.
Our scheme counters this misbehavior with probabilistic checking of signatures (\cref{subsec:coop}) and by cross-checking verifiers (\cref{subsec:detect}).
$SIG$ and $SELF$ verifiers (assuming successful local verification of a said beacon)  are the definitive verifiers, used to compare with $COOP$ or $MAC$ verifiers.
When any conflict exists, corresponding $COOP$ verifier providers are added to \ac{PRL} and corresponding key chain owners are added to \ac{KRL} respectively.
In our scheme, nodes share not only positive $COOP$ verifiers, but negative verifiers too.
With positive verifiers alone, malicious nodes would be detected through signature probabilistic checking.
By introducing negative $COOP$ verifiers, verifiers with opposite validities from benign and malicious nodes can cause conflict, thus makes the receiver checks signature and this way evict the lying malicious node(s).
The two countermeasures can detect and evict malicious nodes effectively and minimize the amount of falsely accepted beacons, thwarting malicious node capability.
We provide an extensive evaluation of misbehavior detection in~\cref{sec:simulation} based on simulation.

Once any falsely validated bogus message is detected, corresponding malicious nodes should be evicted to prevent any further attempt on bogus message validation.
This is established on non-repudiation of such misbehavior by corresponding malicious nodes, which is proven below. We first define \Cref{lemma:1,lemma:2} to prove \Cref{theorem:2}.

\begin{lemma}
	\label{lemma:1}
	Let $B_j$ be an authentic beacon that piggybacks a self-chained verifier that validates an earlier beacon $B_i$ sent from the same node (i.e., the same \ac{PC}) and $i<j$. If the hash function is secure, the successful signature verification on $B_j$ proves authenticity of both $B_j$ and $B_i$, and preserves non-repudiation of disseminating $B_i$.
\end{lemma}

\begin{proof}
	A successful signature verification on $B_j$ is a proof of authentication of $B_i$ (including its self-chained verifiers) by the sender.
	The sender cannot deny the dissemination of $B_i$, because the existence of a third beacon matches the hash value of $B_i$ is practically impossible.
	Similarly, such authentication is transitive to earlier beacons that match self-chained verifiers in $B_i$.
\end{proof}

\begin{lemma}
	\label{lemma:2}
	If a beacon, $B_i$, corresponds to a self-chained verifier or a cooperative verifier in an already verified beacon based on the signature, the sender of the latter cannot deny the validation of $B_i$.
\end{lemma}

\begin{proof}
	A beacon could be accepted if a self-chained verifier or a cooperative verifier piggybacked on a verified beacon matches the hash value of the former beacon.
	In order to deny such a validation, there should exist another beacon, $B_i^\prime$ that matches the same verifier.
	More specifically, if $B_i$ is bogus, there should exist an authentic beacon, which has the same hash value, in order to successfully deny the attempt to validate the bogus beacon, $B_i$.
	Such a hash collision is impossible with a secure hash function, especially considering the time limitation to find such a hash collision, due to ephemeral nature of safety messages.
	Therefore, the self-chained verifier or the cooperative verifier must have been computed based on $B_i$.
\end{proof}

\begin{theorem}
	\label{theorem:2}
	Any successful message validation can be always traced back to the validation by a legitimate node in a non-repudiable manner.
\end{theorem}

\begin{proof}
	It is straightforward that strict signature verification on event messages ensures non-repudiation.
	Conditional anonymity provided by the underlying \ac{VPKI} holds nodes accountable for their messages.
	Similarly, any beacon verified based on the sender's signature provides non-repudiation and ensures the accountability of the sender.
	
	We proved in~\Cref{lemma:1,lemma:2} that validation of a beacon based on a self-chained verifier or a cooperative verifier can also be traced back to a legitimate node in a non-repudiable manner.
	This is especially important for accountability (and follow-up eviction) of malicious nodes attempting to validate bogus beacons based on self-chained verifiers or cooperative verifiers.
\end{proof}

\subsection{DoS-resilience}

For safety beaconing, the challenge is two-fold: efficient neighboring node discovery and keeping track of the discovered nodes; both achieved in terms of beacon reception and verification.
We provide an analysis of the \ac{DoS}-resilience of our scheme before a thorough simulation-based quantitative evaluation in \Cref{sec:simulation}.

Node discovery is facilitated by: 1) continuously queued beacon verification (\cref{subsec:verification}), and 2) cooperative verification (\cref{subsec:coop}).
As in a traditional scheme, each node continuously verifies beacons from the queue for node discovery. 
However, merely verifying beacons from the queue does not guarantee timely discovery of new nodes, because the majority of beacons in the queue could be bogus beacons, when bogus beacon rate overwhelms benign beacon rate.
For more efficient and targeted node discovery, every time a beacon passed any of the three validations, i.e., signature verification, self-chained verification and MAC verification, the positive cooperative verifiers can be used to find non-discovered nodes' beacons.
The beacons found with the positive cooperative verifiers are highly probable to be the benign ones.
The efficient malicious node detection sets a basis for low (non-detected) malicious node ratio.
As a result, extensive cooperative verifiers from honest benign nodes greatly facilitate node discovery.
Even though the cooperative verifiers could be malicious, due to mandatory signature verifications on the found beacons, they will not be accepted and the resultant signature verifications detect misbehaving nodes.

After successful node discovery, one-time keys (i.e., hash chain elements) can be used to keep track of subsequent potentially valid beacons (\cref{subsec:reception}).
Second preimage resistance of hash function ensures that only the key chain owner knows and can disclose the correct one-time keys.
Therefore, even if multiple (single-hop) beacons, piggybacking the correct \{one-time key, \ac{PC}\} pairs, are received during each time interval, only the first of them needs to be queued; the rest can be dropped, because they are sent after overhearing the original beacon.
In case the original beacon was not received due to packet loss, a received masqueraded beacon may be considered as the potentially valid one.
However, the effect of such bogus beacons is limited, because at most one bogus beacon is queued for each time slot for each \ac{PC}: a significant improvement over the traditional scheme, with which all bogus beacon signatures need to be verified (or until the valid one is verified for that time slot) before they can be dropped.
Moreover, negative cooperative verifiers and self-chained verifications can help eliminate such masqueraded beacons.

We explicitly assign CPU time ratios for discovered/non-discovered nodes' beacon verification: $Ratio_{D}$ of time for discovered nodes' beacons and $Ratio_{ND}$ for node discovery, while $Ratio_{D} + Ratio_{ND} = 1$.
The checks on key chain help to filter out bogus beacons attached the discovered \acp{PC}, and the explicitly assigned CPU time ratio enable receivers to verify their discovered nodes' beacons, even if attackers flood and overwhelm receiver queues with bogus beacons attached random signatures and \acp{PC} (thus non-discovered nodes' beacons, from the perspective of receivers).

Event message verification is established on beacon verification.
The \ac{DoS}-resilient beacon verification ensure timely caching of event message facilitator before the corresponding event message dissemination.
Nodes efficiently capture and queue benign event messages matching the cached facilitators and drop bogus event messages.

%% file: section_compsec/simulation.tex
\section{Quantitative Evaluation}
\label{sec:simulation}

We present detailed simulation results with realistic mobility and communication models, and processing delays.
We show that our scheme provides timely message validation with low beacon expiration ratio and low node discovery delay even under loaded networks or \ac{DoS} attacks than the baseline scheme (the prior approaches that strictly verify every message signature).\footnote{We do not compare with our earlier work~\cite{jin2019resilient}, because, as explained earlier, the adopted symmetric key based authentication does not provide non-repudiation and accountability.}
Considering internal malicious nodes that abuse features of the scheme, notably the cooperative verification, we show our scheme is resilient to false validations and can effectively detect malicious nodes.
With an evaluation of event-driven message dissemination, we show our scheme can facilitate reception and validation of various message types.

\subsection{Simulation Setting}

\begin{table}[ht]
	\centering
	\begin{minipage}[b]{0.4\textwidth}
		\centering
		\includegraphics[width=0.8\columnwidth]{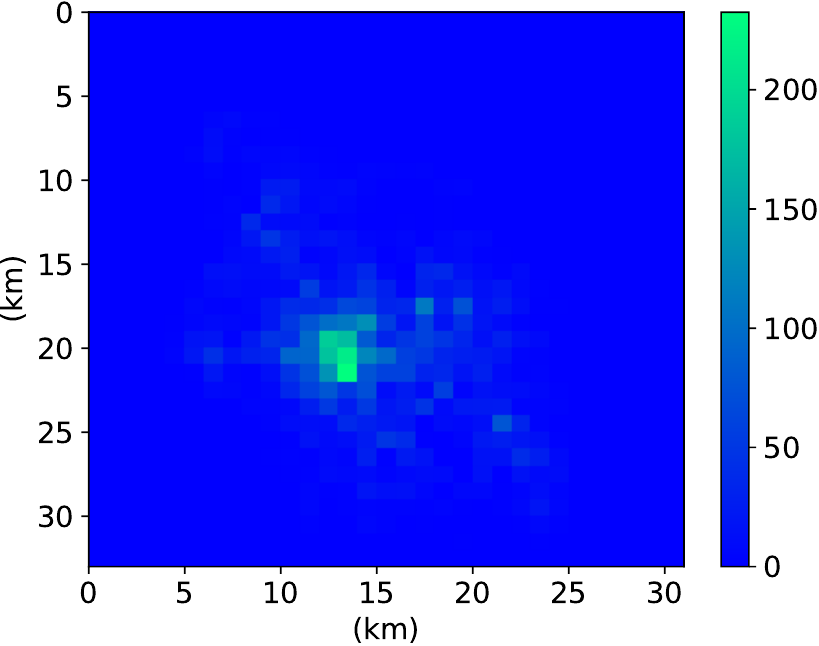}
		\captionof{figure}{Node density in TAPASCologne scenario at 12:30 pm.}
		\label{fig:density}
	\end{minipage}\hfill
	\begin{minipage}[b]{0.4\textwidth}
		\centering
		\scriptsize
		\begin{tabular}{ l | c }
			\hline \hline
			Bitrate & \textbf{\emph{6}}, 27 $Mbps$ \\\hline
			$Pr_{check}$ & 0, \emph{\textbf{0.2}}, 0.5, 0.8 \\\hline
			$\alpha$ & 0, 1, 2, \textbf{\emph{3}}, 4 \\\hline
			$k$ & \textbf{\emph{3}} \\\hline
			$T_{blife}$ & \textbf{\emph{1}} $sec$ \\\hline
			$\tau$  & 0.4, 2, \textbf{\emph{4}} $msec$ \\\hline
			$\gamma$ & \textbf{\emph{10}} $Hz$ \\\hline
			$\beta_1$ & \textbf{\emph{1}}, 2 \\\hline
			$\beta_2$ & 2, \textbf{\emph{3}} \\\hline
			$\gamma_{DoS}$ & \textbf{\emph{250}}, 500, 1000 $Hz$ \\\hline
			$Ratio_{adv}$ & 0.1, 0.3, \emph{\textbf{0.5}} \\\hline
			$\{Ratio_{S}, Ratio_{V}\}$ &  \emph{\textbf{\{0.5, 0.5\}}} \\\hline
			$\{Ratio_{D}, Ratio_{ND}\}$ & \emph{\textbf{\{0.5, 0.5\}}} \\\hline
			\hline
		\end{tabular}
		\vspace{1em}
		\captionsetup{skip=0.5pt}
		\captionof{table}{System Parameters (\textbf{\emph{Bold}} for Default Setting)}
		\label{table:parameter}
	\end{minipage}\hfill
	
\end{table}

We use OMNeT++~\cite{varga2008overview} for a packet-level simulation with IEEE 802.11p module for \ac{V2V} communication provided by Veins~\cite{sommer2011bidirectionally}.
We consider a maximum communication range of $200m$ with a default 6 Mbps bit rate.
SUMO provides a microscopic mobility simulation module that serves Veins with mobility traces. 
We use the SUMO~\cite{SUMO2012} TAPASCologne scenario~\cite{uppoor2013generation} to simulate vehicle mobility, with a penetration ratio of 50\%, i.e., 50\% of nodes disseminate safety messages and participate in our scheme.
We consider a base value for processing power and thus $\tau$, that reflects recent literature~\cite{calandriello2011performance,petit2013authentication,baee2019broadcast,pan2019secure,preserved32,campvsc5}.
Then, we increase processing power by an order of magnitude, to the minimum $\tau$ latest OBUs claim to support.

We simulate with vehicle traces from 12:00 pm, and vehicles start safety beaconing from 12:30 pm and continue for 2 minutes before simulations conclude: the first 30 minutes are used for filling up the network with vehicles.
\Cref{fig:density} shows a heat map of node density at 12:30 pm.
We use a $2\ km \times 2\ km$ area ($Region_{sim}$) with the densest node population of the city.
Nodes start safety beaconing within the central $1\ km \times 1\ km$ area ($Region_{beacon}$) of the simulated area, and results are collected from nodes within the central $0.5\ km \times 0.5\ km$ ($Region_{result}$) area.
Although vehicles, in a real-world scenario, keep beaconing without any region restriction, we choose such an area, although much smaller than the full city size, in order to keep the simulations manageable, while being able to capture performance in the most significant area.
Internal adversaries start their attacks when they enter the central $0.5\ km \times 0.5\ km$ area ($Region_{attack}$), affecting more nodes in the denser area.\footnote{If the adversaries start attacking in $Region_{beacon}$, they can be detected more easily with signature verifications, because less nodes exist at the border of $Region_{beacon}$, thus less beacon arrivals.}
16 bogus beacon generators (i.e., \ac{DoS} attackers), which form a $4 \times 4$ matrix, are placed at the center $0.6\ km \times 0.6\ km$ area.
Nodes start disseminating safety beacons at 12:30 pm.
This captures a situation that vehicles change their \acp{PC} simultaneously, as per the privacy-preserving \ac{PC} changing policy~\cite{khodaei2018secmace}, which is the most challenging scenario in terms of node discovery.
As a result, some nodes will be attacked by more than one bogus beacon generators.
\Cref{table:parameter} shows simulation parameters.
We consider an average $\tau$ that includes both signature verification and relevant scheme operations (e.g., hash computation and string comparison).
The former could be generally one or two order(s) of magnitude more computationally expensive than the latter one.
We consider a beacon size to be 300 bytes, and each hash digest to be 20 bytes (i.e., SHA-1 hash size).
Results are averaged over five randomly seeded runs for each simulation setting.

\subsection{Resilience to DoS Attacks}

We evaluate different beacon validation metrics: \textbf{waiting time}, \textbf{expiration ratio}, \textbf{validation type ratio} and \textbf{node discovery delay}. \emph{Waiting time} is defined as the time the beacon stays in queue until its verification.
\emph{Expiration ratio} is the ratio of beacons that expired before any validation.
\emph{Validation type ratio} shows percentage of beacon validation based on each method. Another important criterion for availability of the scheme is short discovery delay of new neighboring nodes.
The continuous awareness of a neighboring node is only possible if beacons from the node can be continuously verified and accepted.
With the ability to efficiently track potentially valid subsequent beacons based on hash chains, the challenge lies in verification of a beacon from the new neighboring node so that the corresponding hash chain can be trusted.
We evaluate this criteria with \emph{node discovery delay}: delay between the reception of the first valid beacon from a new neighboring node and the verification of at least one (possibly later) beacon from that same node.

\begin{figure}[tp]
	\centering
	\begin{subfigure}[b]{.24\textwidth}
		\includegraphics[width=\columnwidth]{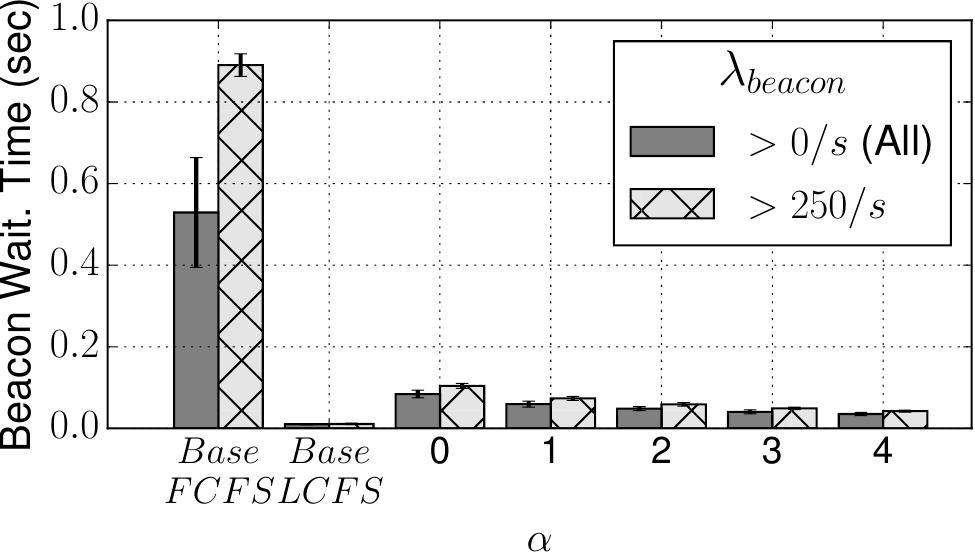}
		\caption{}
		\label{subfig_wait_num}
	\end{subfigure}\hfill
	\begin{subfigure}[b]{.24\textwidth}
		\includegraphics[width=\columnwidth]{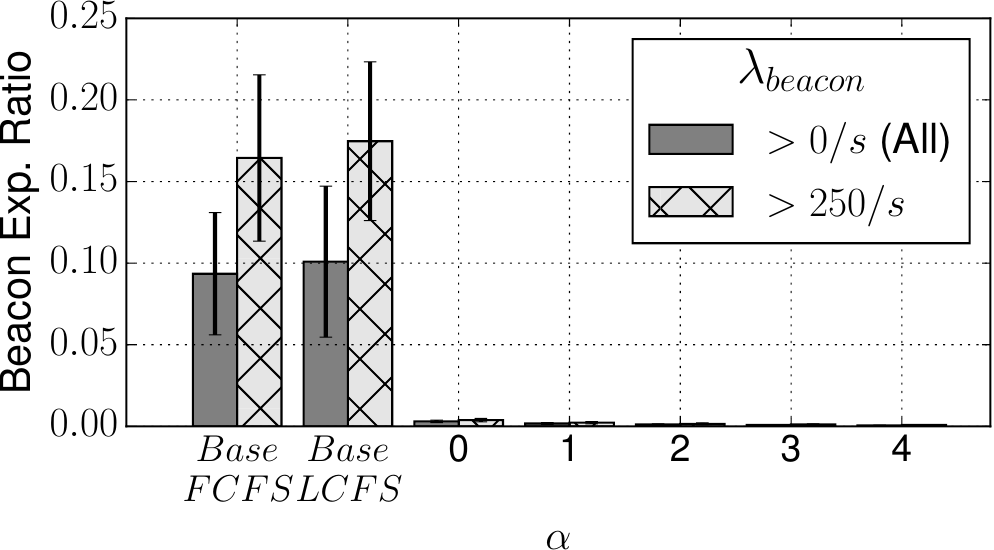}
		\caption{}
		\label{subfig_expire_num}
	\end{subfigure}\hfil
	\caption{Beacon validation metrics as a function of $\alpha$ when in benign network.}
	\label{fig_benign_num}
\end{figure}
\begin{figure}
	\begin{subfigure}[b]{.24\textwidth}
		\includegraphics[width=\columnwidth]{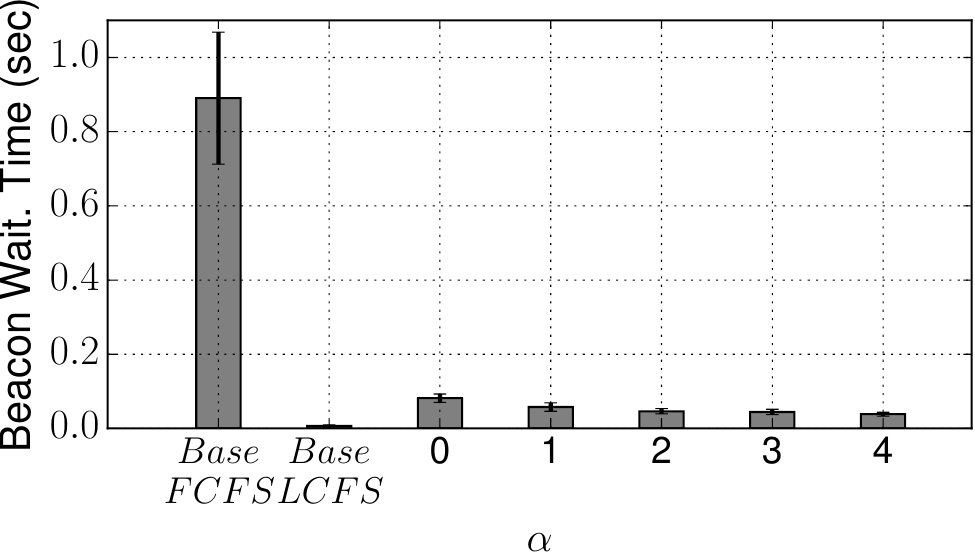}
		\caption{}
		\label{subfig_wait_dos_num}
	\end{subfigure}\hfill
	\begin{subfigure}[b]{.24\textwidth}
		\includegraphics[width=\columnwidth]{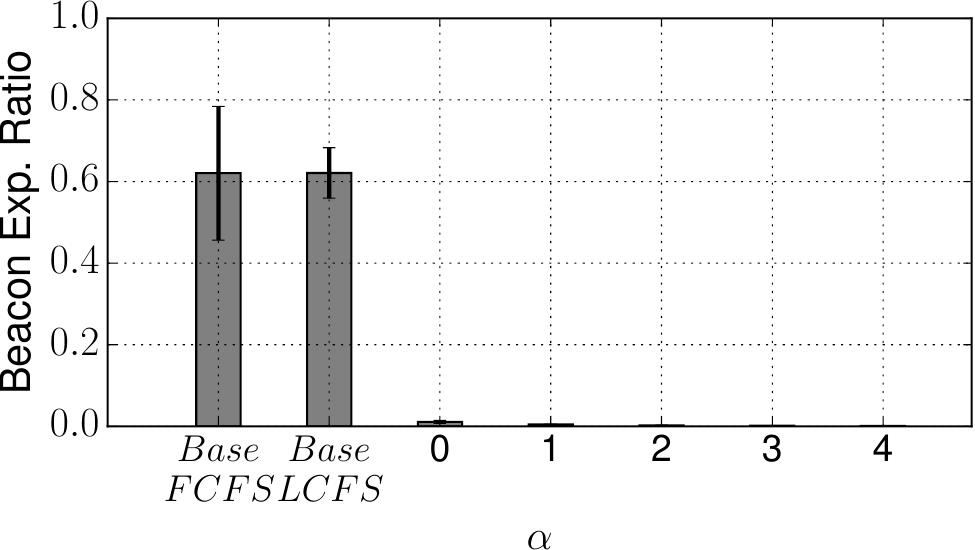}
		\caption{}
		\label{subfig_expire_dos_num}
	\end{subfigure}\hfill
	\begin{subfigure}[b]{.24\textwidth}
		\includegraphics[width=\columnwidth]{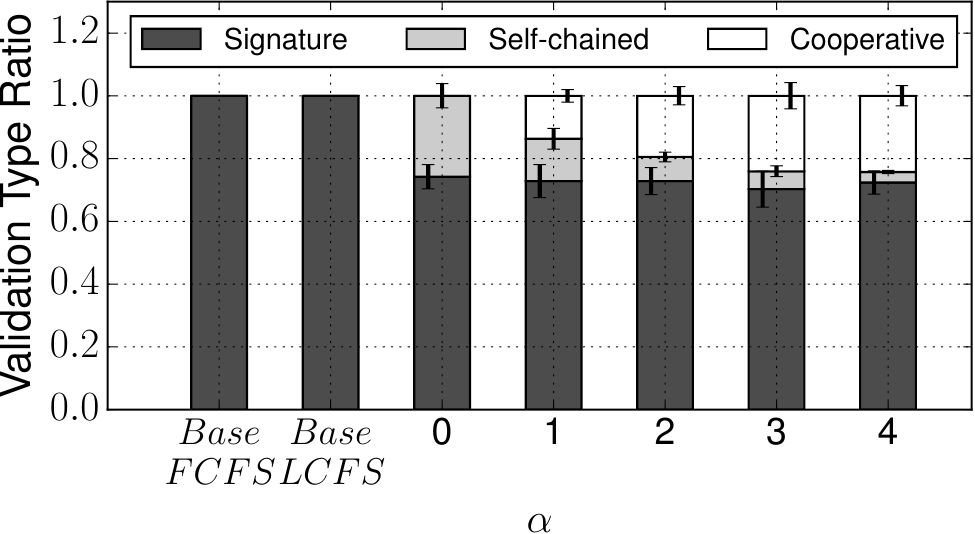}
		\caption{}
		\label{subfig_type_dos_num}
	\end{subfigure}\hfill
	\begin{subfigure}[b]{.24\textwidth}
		\includegraphics[width=\columnwidth]{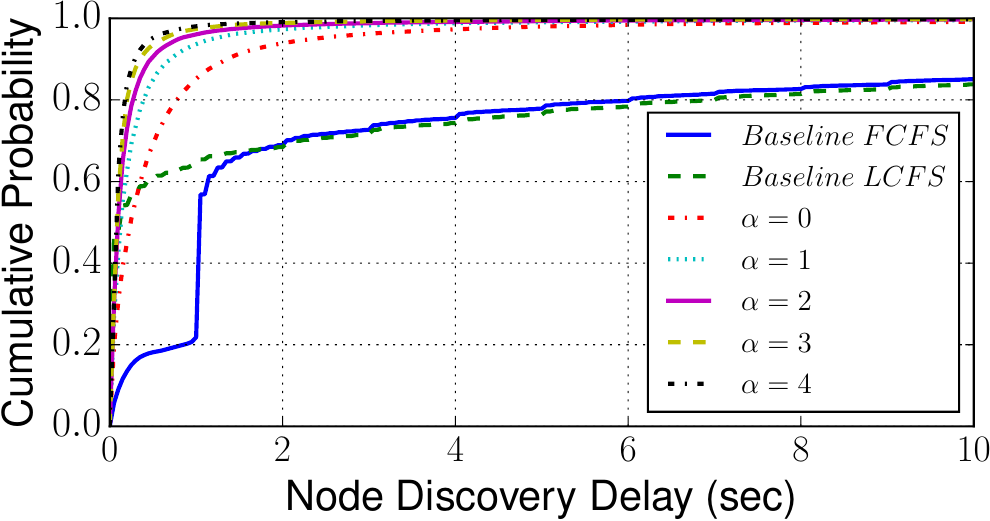}
		\caption{}
		\label{subfig_delay_dos_num}
	\end{subfigure}
	\caption{Beacon validation metrics as a function of $\alpha$ when under the DoS attack.}
	\label{fig_dos_num}
\end{figure}

We evaluate the baseline scheme and our scheme in benign and \ac{DoS} scenarios.
The baseline scheme, with none of the \ac{DoS}-resilient features, nodes verify beacons independently with pure (i.e., not semi-randomized) \ac{FCFS} or \ac{LCFS} queue processing.
\Cref{fig_benign_num,fig_dos_num} show the metrics for the baseline scheme (referred as $Base$ in the figures) and our scheme.
For the benign scenario (\cref{fig_benign_num}), apart from average beacon waiting time and expiration ratio for all nodes ($\lambda_{beacon} > 0 /s$.), we evaluate the metrics for nodes that $\lambda_{beacon}$ is higher than $1/\tau=250/s$ (i.e., higher message arrival rate than a queue can sustain).
This essentially captures the performance of our scheme in benign networks when nodes are overloaded with benign beacons, thus evaluates the scalability provided by our scheme.
For the benign scenarios, we see the \ac{FCFS} baseline scheme exhibits high beacon waiting time while the \ac{LCFS} baseline scheme provides low waiting time: the former approach verifies the earliest beacon in the queue, while the latter approach always verifies the latest beacon.
Due to the consistent $\tau$, the amount of verifiable beacons per time unit are roughly the same for the both approaches.
Therefore, the two approaches exhibit similar expiration ratios.
As expected, the baseline scheme exhibits high beacon expiration ratios: \cref{subfig_expire_num} shows around 10\% and 16\% of beacons from benign nodes have to be dropped due to expiration, for $\lambda_{beacon} > 0 /s$ and $\lambda_{beacon} > 250 /s$ respectively.
When our scheme is adopted, expiration ratios significantly decreases, while maintaining reasonably low beacon waiting times (see \cref{subfig_wait_num,subfig_expire_num} with $\alpha \geq 0$).
Higher $\alpha$ introduces higher communication overhead, but nodes become more scalable and resilient to both \ac{DoS} attacks and malicious nodes, while such improvement becomes moderate as $\alpha$ increases.

For the \ac{DoS} scenarios (\cref{fig_dos_num}), the \ac{LCFS} baseline scheme still provides very low waiting time and the \ac{FCFS} baseline scheme provides an average waiting time close to 1s, i.e., the beacon lifetime, because the queues are always loaded with high-rate beacons and the verified beacons almost reached the end of their lifetimes. The expiration ratios increases to around 60\% due to computation power wasted on bogus beacon verification. Thanks to continuous (potentially) valid beacon tracking leveraging hash chains and explicit time allocation for discovered and non-discovered nodes, our scheme guarantees low beacon expiration ratios even under the DoS attacks: roughly same as those for the benign scenario (see \cref{subfig_expire_num,subfig_expire_dos_num} with $\alpha \geq 0$).

\Cref{subfig_type_dos_num} shows ratios of verified beacons based on each validation methods. With higher $\alpha$, higher ratios of beacons are validated based on $COOP$ verifiers, while lower ratios of beacons need to wait for self-chained verifications. This results in lower average waiting time as $\alpha$ increases (\cref{subfig_wait_num,subfig_wait_dos_num}): $COOP$ verifiers can validate fresher beacons while $SELF$ verifiers validate beacons that are received several beacon intervals earlier.

\Cref{subfig_delay_dos_num} shows \acp{CDF} for node discovery delay. For the baseline scheme, each node independently discovers neighbors and beacon arrival rate is higher than beacon verification rate for some nodes in the simulations, which result in higher node discovery delays.
There is a significant improvement even with $\alpha = 0$ thanks to expedited queue processing based on self-chained verifications.
There is slight improvement with $\alpha > 0$ than with $\alpha = 0$, while the lines almost overlap with higher positive $\alpha$ values.
Improvements with positive $\alpha$ values (i.e., $\alpha > 0$) are still observable, while the lines almost overlap with higher positive $\alpha$ values.
When $\alpha = 0$, around 77\% of nodes can be discovered within $0.5s$, and increases to around 90\% when $\alpha = 1$.
It improves further, but only moderately, with higher $\alpha$ values, e.g., around 96\% and 97\% of nodes are discovered within $0.5s$ when $\alpha$ = 3 and 4 respectively.

\begin{figure*}[t]	
	\begin{subfigure}[b]{.24\textwidth}
		\includegraphics[width=\columnwidth]{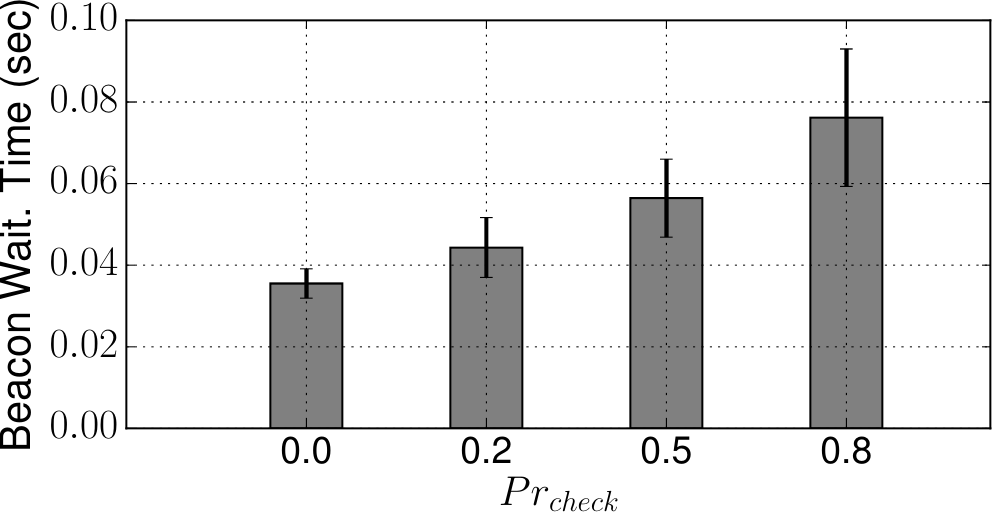}
		\caption{}
		\label{subfig_wait_dos_checkProb}
	\end{subfigure}\hfill
	\begin{subfigure}[b]{.24\textwidth}
		\includegraphics[width=\columnwidth]{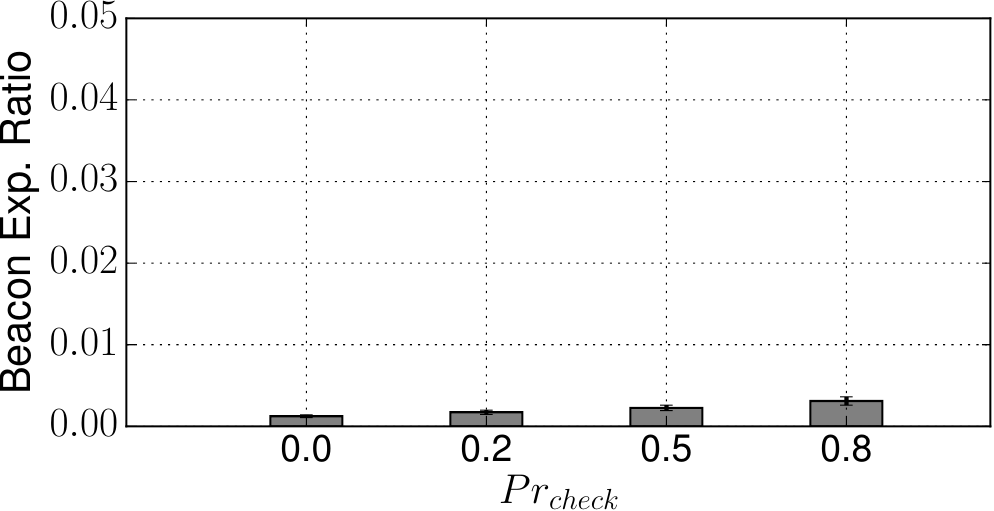}
		\caption{}
		\label{subfig_expire_dos_checkProb}
	\end{subfigure}\hfill
	\begin{subfigure}[b]{.24\textwidth}
		\includegraphics[width=\columnwidth]{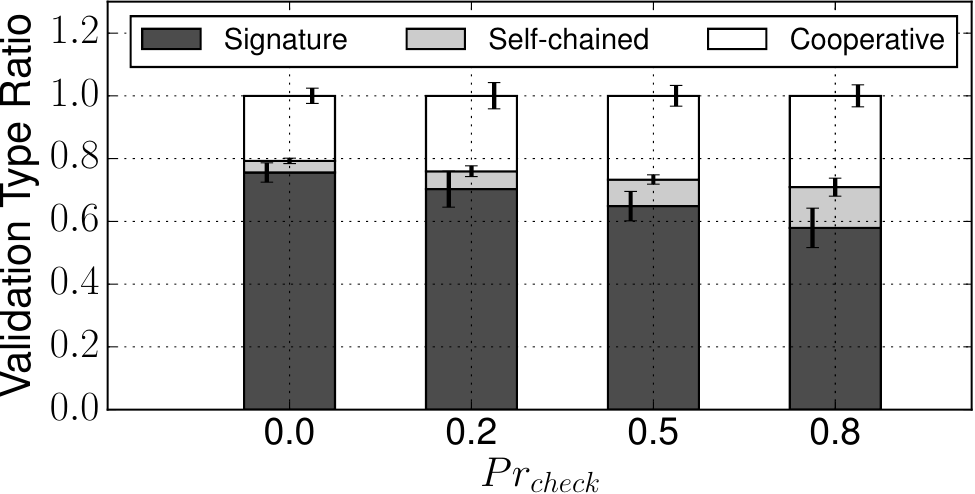}
		\caption{}
		\label{subfig_type_dos_checkProb}
	\end{subfigure}\hfill
	\begin{subfigure}[b]{.24\textwidth}
		\includegraphics[width=\columnwidth]{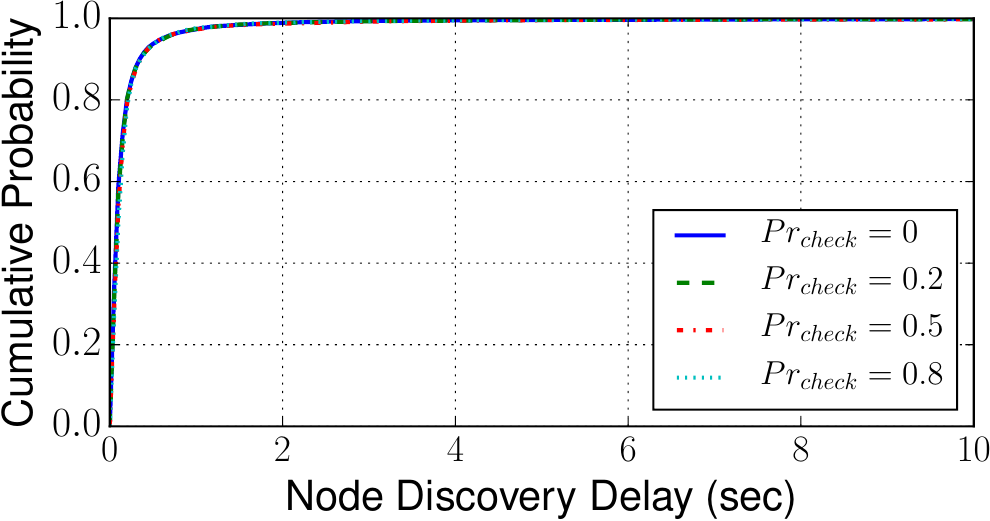}
		\caption{}
		\label{subfig_delay_dos_checkProb}
	\end{subfigure}
	\caption{Beacon validation metrics as a function of $Pr_{check}$ under the DoS attack. Default: $\alpha = 3$.}
	\label{fig_dos_checkprob}
\end{figure*}

We show beacon validation metrics as a function of $Pr_{check}$ (\cref{fig_dos_checkprob}).
In general, our scheme still maintains reasonably low waiting time, expiration ratio and node discovery delay (\cref{subfig_wait_dos_checkProb,subfig_expire_dos_checkProb,subfig_delay_dos_checkProb}).
In the evaluation, we categorize beacons that probabilistically checked into cooperatively verified.
With higher $Pr_{check}$, less beacons can be verified based on signatures (\cref{subfig_type_dos_checkProb}), because more cooperatively validated beacons need to be checked, which results in more beacons validated based on the alternative methods.

\begin{figure*}
	\centering
	\begin{subfigure}[b]{.33\textwidth}
		\includegraphics[width=\columnwidth]{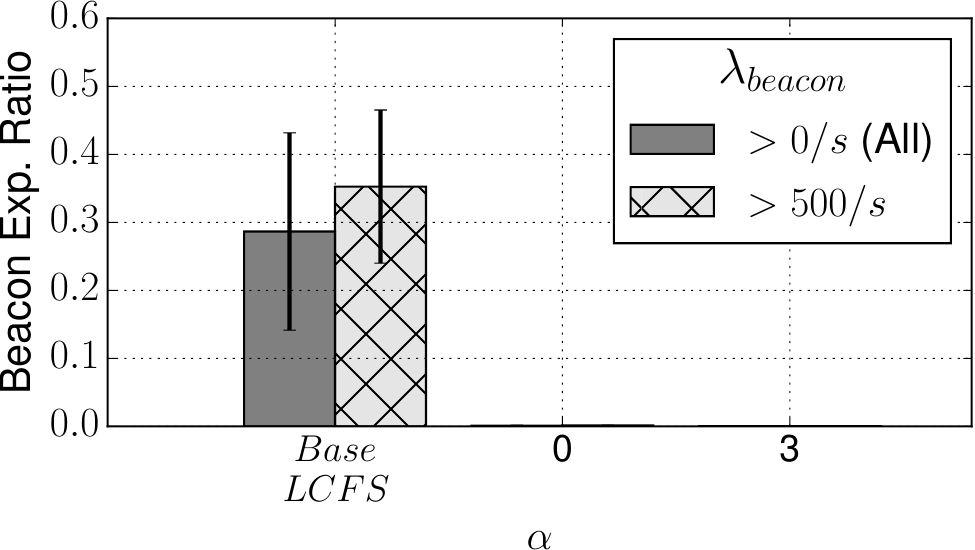}
		\caption{}
		\label{subfig_expire_dos_2_num}
	\end{subfigure}\hfill
	\begin{subfigure}[b]{.33\textwidth}
		\includegraphics[width=\columnwidth]{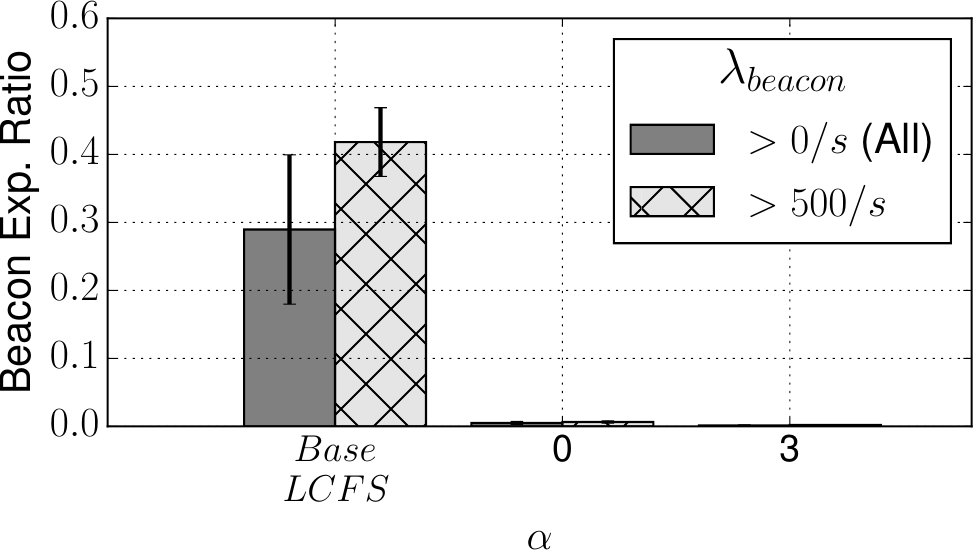}
		\caption{}
		\label{subfig_expire_dos0002_2_num}
	\end{subfigure}\hfill
	\begin{subfigure}[b]{.33\textwidth}
		\includegraphics[width=\columnwidth]{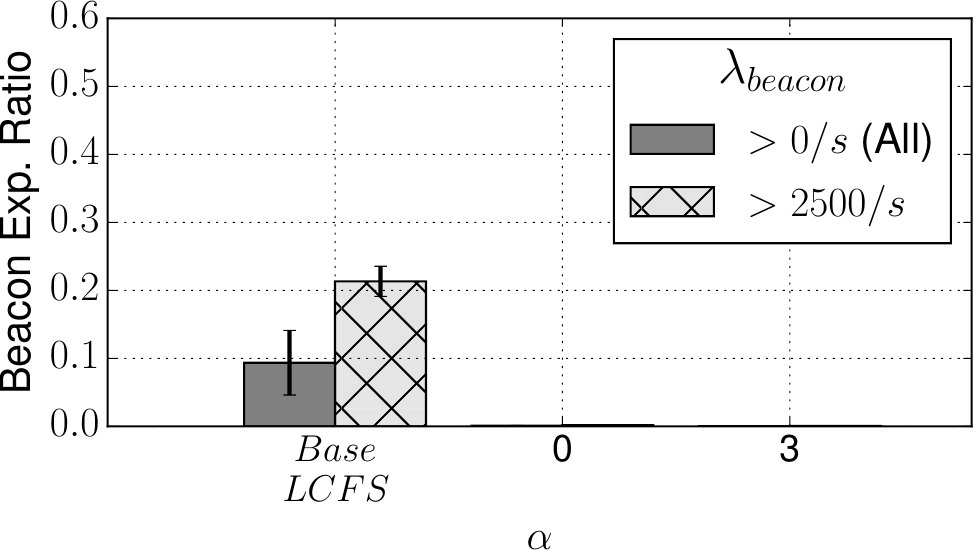}
		\caption{}
		\label{subfig_expire_dos1000_recv_04_num_max}
	\end{subfigure}\hfill
	\begin{subfigure}[b]{.33\textwidth}
		\includegraphics[width=\columnwidth]{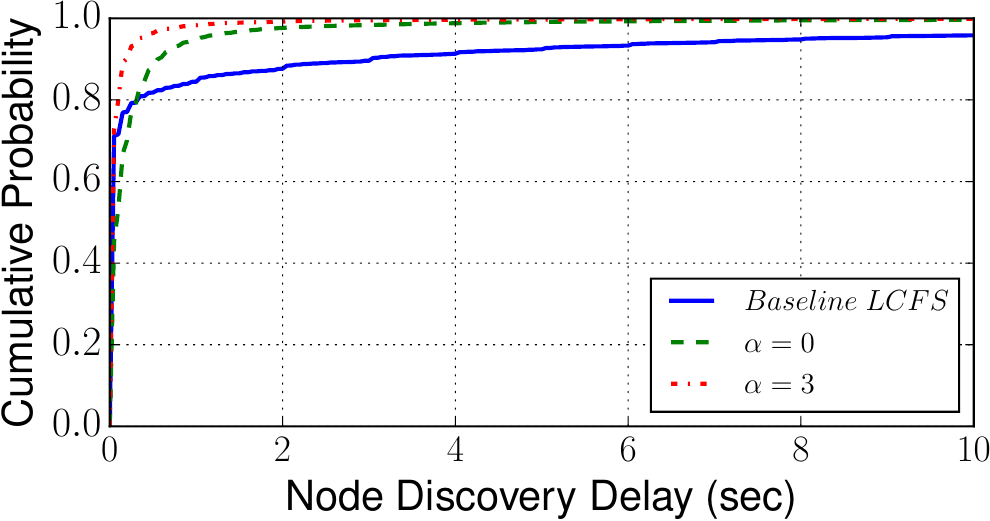}
		\caption{}
		\label{subfig_delay_dos_2_num}
	\end{subfigure}\hfill
	\begin{subfigure}[b]{.33\textwidth}
		\includegraphics[width=\columnwidth]{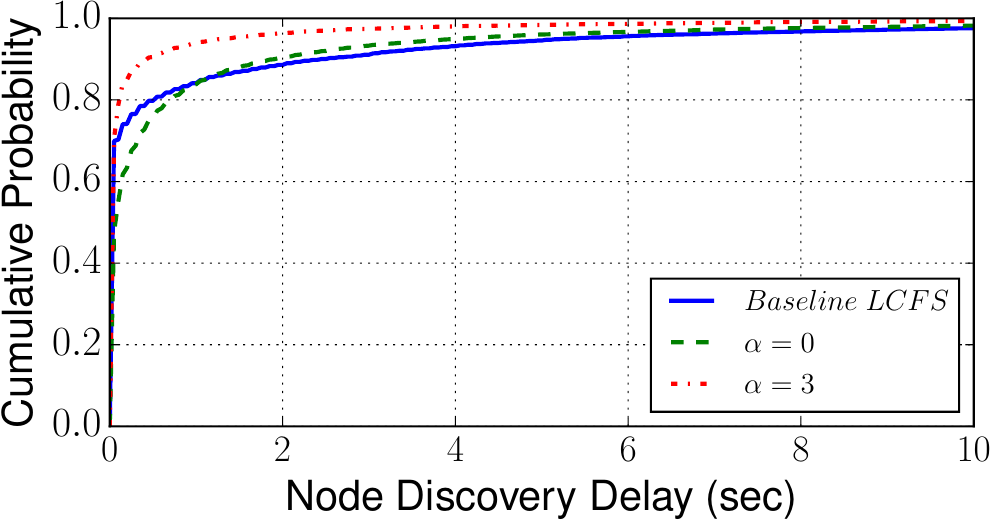}
		\caption{}
		\label{subfig_delay_dos0002_2_num}
	\end{subfigure}\hfill
	\begin{subfigure}[b]{.33\textwidth}
		\includegraphics[width=\columnwidth]{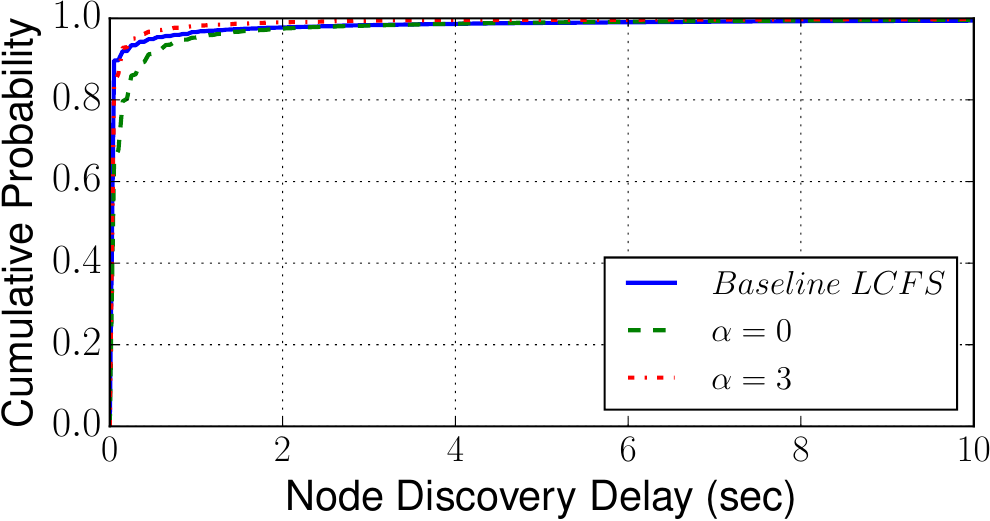}
		\caption{}
		\label{subfig_delay_dos1000_recv_04_num_max}
	\end{subfigure}\hfill
	\caption{Beacon expiration ratio and node discovery delay as a function of $\alpha$ when under a clogging DoS attack. (\subref{subfig_expire_dos_2_num}), (\subref{subfig_delay_dos_2_num}) Bitrate = 6 Mbps, $\tau = 2ms$ and $\gamma_{DoS} = 250\ Hz$. (\subref{subfig_expire_dos0002_2_num}), (\subref{subfig_delay_dos0002_2_num}) Bitrate = 6 Mbps, $\tau = 2ms$ and $\gamma_{DoS} = 500\ Hz$. (\subref{subfig_expire_dos1000_recv_04_num_max}), (\subref{subfig_delay_dos1000_recv_04_num_max}) Bitrate = 27 Mbps, $\tau = 0.4ms$ and $\gamma_{DoS} = 1000\ Hz$.}
	\label{fig_dos_rate}
\end{figure*}

We continue evaluation with lower message verification delay and higher bogus beacon rates (\cref{fig_dos_rate}). With a lower $\tau = 2\ msec$ (\cref{subfig_expire_dos_2_num,subfig_expire_dos0002_2_num,subfig_delay_dos_2_num,subfig_delay_dos0002_2_num}), beacon expiration ratios are still high for the baseline scheme and significantly lower with our scheme, and node discovery delays improve with higher $\alpha$ values.
For example, when $\tau = 2\ msec$ and $\gamma_{DoS} = 500\ Hz$ (\cref{subfig_delay_dos0002_2_num}), only around 59\% of nodes can be discovered within $0.5\ sec$ with $\alpha = 0$, but the value significantly increases to 88\% with $\alpha = 3$.
Moreover, the \ac{CDF} converges towards 100\% much faster than the baseline scheme.
Next, we evaluate our scheme considering high-end vehicular \acp{OBU}.
We decrease $\tau$ by one order of magnitude from the default value, thus $\tau = 0.4\ msec$, and set the bitrate of each node to the maximum 27 Mbps~\cite{teixeira2014vehicular,sepulcre20176}.
With this setting, we see generally improved results even under higher bogus beacon rates, $\gamma_{DoS} = 1000\ Hz$, due to one order of magnitude shorter beacon processing delays.
With the baseline scheme, around 95\% of nodes can be discovered within 0.5s, while the values are 91\% and 97\% for our scheme with $\alpha =$ 0 and 3 respectively.
Our scheme slightly outperforms the baseline scheme only with a positive $\alpha$ value (e.g., 3), but our scheme ensures much lower expiration ratios (almost 0\%), compared to around 10\% and 20\% of expiration ratios for the general and loaded cases respectively.
Although the expiration ratios for the baseline scheme are less significant than those in \cref{subfig_expire_dos_2_num,subfig_expire_dos0002_2_num}, in highly congested networks that already experience high packet loss rates, even the seemingly relatively low expiration ratio, e.g., 20\%, could be critical for safety application functionality.
Similarly, we see reasonably low waiting time values for the three scenarios from the simulation results (not shown due to space limitation).
In general, our scheme provides less significant improvements with lower processing delays, due to the relatively moderate bandwidth increase (i.e., 4.5 times higher) compared to the improved (10 times higher) computational power.
However, we can expect better performance with our scheme when advanced communication technology is used, e.g., 5G with higher bandwidth and better congestion control, with which much higher message rates are expected.

\subsection{Resilience to Malicious Nodes}

\begin{figure*}[h]
	\centering
	
	\begin{subfigure}[b]{.33\textwidth}
		\includegraphics[width=\columnwidth]{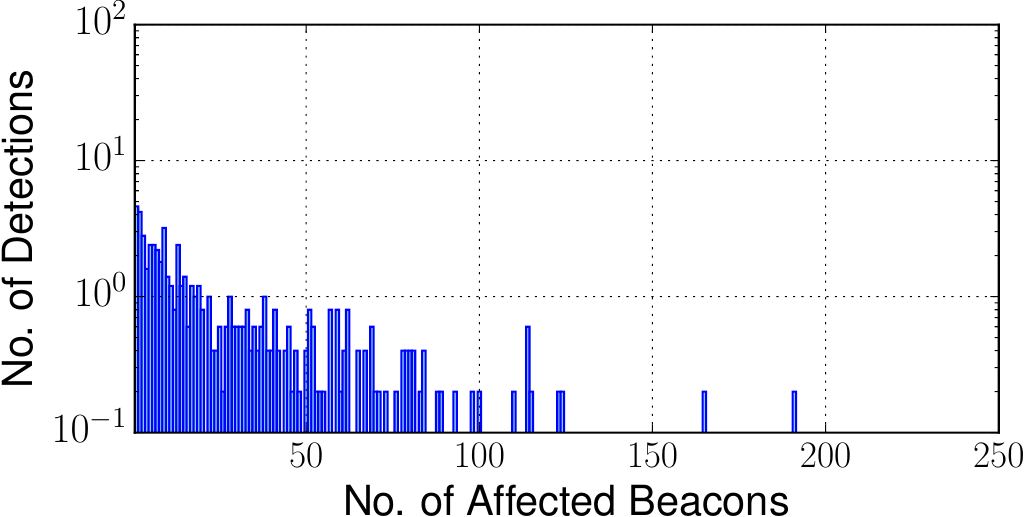}
		\caption{}
		\label{subfig_affect_advNoSpreadNegConflictCheckRatio_0.1}
	\end{subfigure}\hfill
	\begin{subfigure}[b]{.33\textwidth}
		\includegraphics[width=\columnwidth]{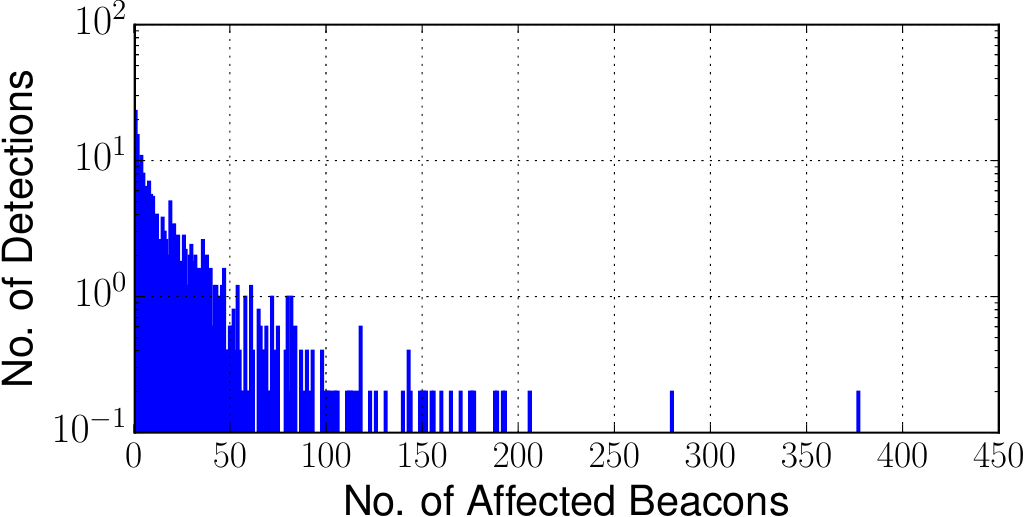}
		\caption{}
		\label{subfig_affect_advNoSpreadNegConflictCheckRatio_0.3}
	\end{subfigure}\hfill
	\begin{subfigure}[b]{.33\textwidth}
		\includegraphics[width=\columnwidth]{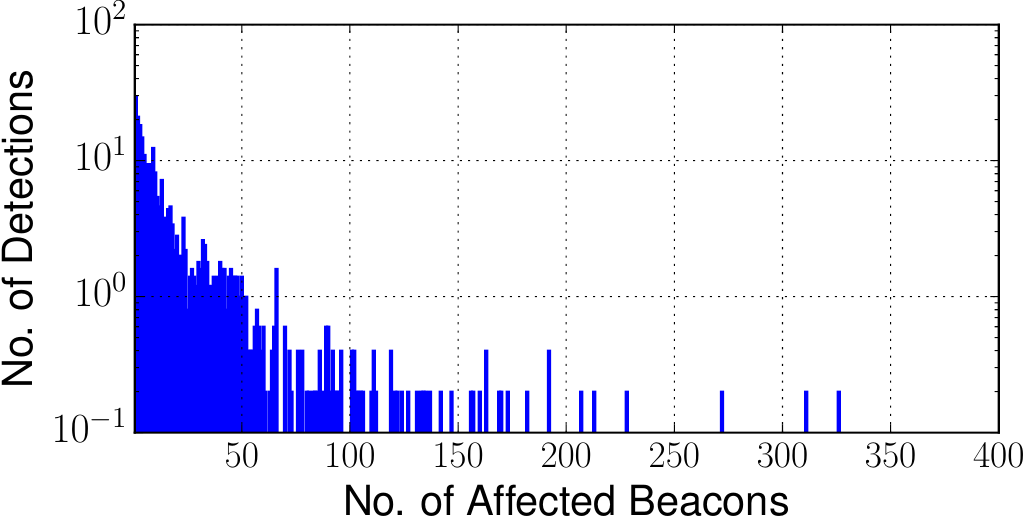}
		\caption{}
		\label{subfig_affect_advNoSpreadNegConflictCheckRatio_0.5}
	\end{subfigure}

	\begin{subfigure}[b]{.33\textwidth}
		\includegraphics[width=\columnwidth]{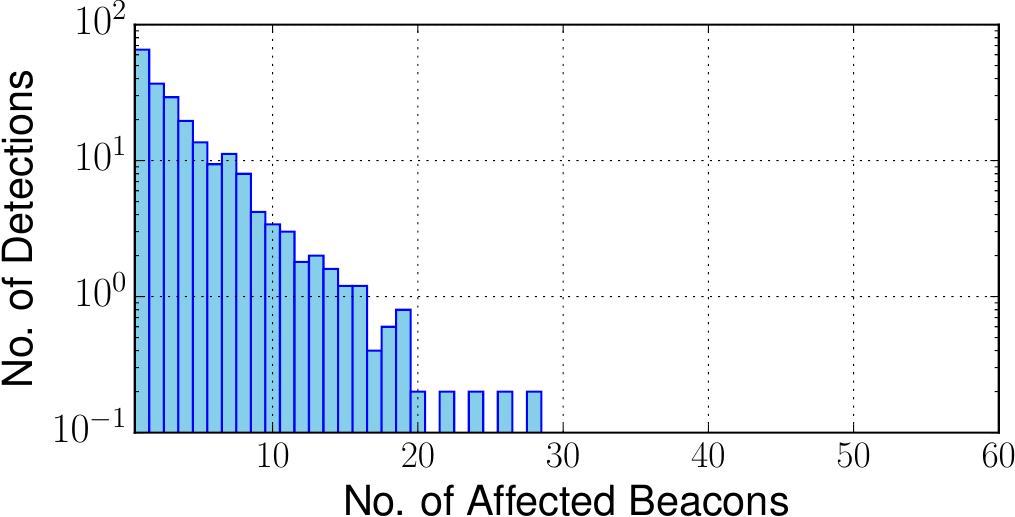}
		\caption{}
		\label{subfig_affect_advNoSpreadNegConflictCheckProb_0.2}
	\end{subfigure}
	\begin{subfigure}[b]{.33\textwidth}
		\includegraphics[width=\columnwidth]{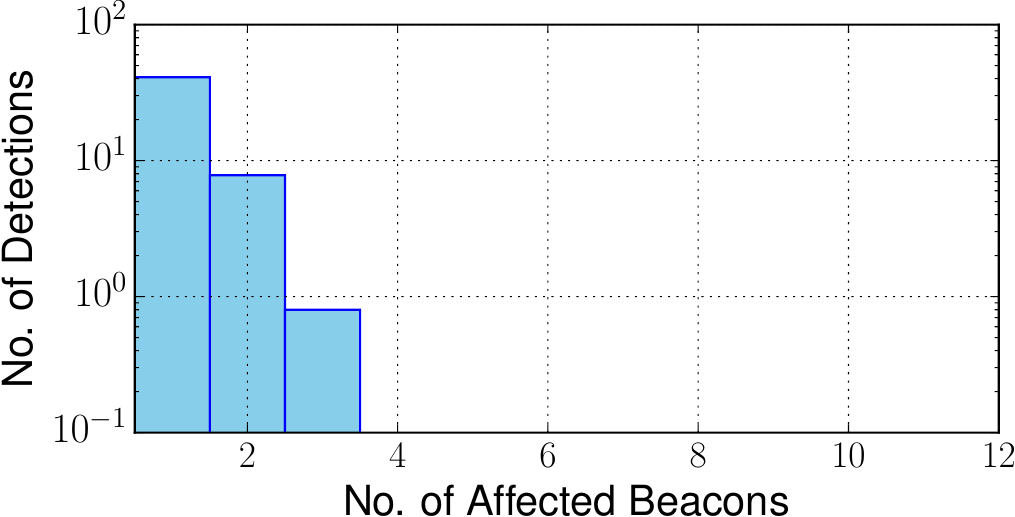}
		\caption{}
		\label{subfig_affect_advNoSpreadCheckProb_0.2}
	\end{subfigure}
	\caption{Histogram of numbers of affected beacons when under the DoS attack and in the presence of malicious nodes. (\subref{subfig_affect_advNoSpreadNegConflictCheckRatio_0.1}),(\subref{subfig_affect_advNoSpreadNegConflictCheckRatio_0.3}),(\subref{subfig_affect_advNoSpreadNegConflictCheckRatio_0.5}) No protection with $Ratio_{adv} = 0.1, 0.3, 0.5$. (\subref{subfig_affect_advNoSpreadNegConflictCheckProb_0.2}) Probabilistic checking only. (\subref{subfig_affect_advNoSpreadCheckProb_0.2}) Both probabilistic checking and cross-checking. (Default: $Ratio_{adv}=0.5$, $Pr_{check} = 0.2$)}
	\label{fig_affect_adv_ConflictCheckProb}
\end{figure*}

We continue with evaluation of resilience to internal malicious nodes, i.e., nodes that falsely validate bogus beacons. The only way to make a benign node (i.e., victim) accepting bogus beacons, is through cooperative verifiers. Moreover, this is only possible after the \ac{PC} on the bogus beacons has been discovered already by the victim and the subsequent bogus beacons are piggybacked with correct one-time keys and MACs. In the simulations, we assume $Ratio_{adv}$ of nodes are malicious, and $Ratio_{S}$ of malicious nodes are malicious senders that disseminate bogus beacons with correct one-time keys and $Ratio_{V}$ of malicious nodes are malicious validators that overhear the malicious bogus beacons and validate the bogus beacons through their own false positive cooperative verifiers. Our scheme has two features for malicious node detection: probabilistic checking and verifier cross-checking. In this evaluation, we record, for each benign and malicious nodes pair, numbers of bogus beacons the malicious (validator) node made the benign node accept, through cooperative verifiers, before the malicious node is detected by the benign node. We term accepted bogus beacon as affected beacon. A benign node could be attacked by multiple malicious validators and a malicious validator could successfully attack multiple benign nodes, but we consider each benign and malicious nodes pair as one instance and count the number of affected beacon for each pair.

\Cref{subfig_affect_advNoSpreadNegConflictCheckRatio_0.1,subfig_affect_advNoSpreadNegConflictCheckRatio_0.3,subfig_affect_advNoSpreadNegConflictCheckRatio_0.5} show histograms of numbers of affected beacon under different $Ratio_{adv}$ values without any detection features.
In this case, detection is possible only when a node verified signature of a cooperatively validated bogus beacon, for node discovery.
A  malicious validator can make a benign node accept more than 100 (more than 300, for the worst case when $Ratio_{adv}= 0.3\ and \ 0.5$) bogus beacons.
With such amount of bogus beacons accepted by a benign node, vehicle operations and even vehicle safety can be heavily affected.
Next, we evaluate scenarios with probabilistic checking only.
With a positive $Pr_{check} = 0.2$, (\cref{subfig_affect_advNoSpreadNegConflictCheckProb_0.2}), we see the amount of affected beacons significantly decrease.
Although this shows the importance of probabilistic checking for malicious node detection, a benign node could still falsely accept around 20 bogus beacons validated by a malicious node.
We can expect the number should decrease with higher $Pr_{check}$ value, but it would incur higher computation overhead, conflicting with the goal of our scheme.
We evaluate the performance with the adoption of both detection features (\cref{subfig_affect_advNoSpreadCheckProb_0.2}).
The results significantly improves due to higher chances for misbehavior detection based on cross-checking (see \Cref{sec:analysis}).

We see from the results that only few bogus beacons are accepted while an overwhelming majority of benign beacons can still provide redundancy to tolerate such bogus beacon consumption. The only way to disrupt the system is still providing authenticated false data. We see from the results that our scheme does not degrade security, while significantly improves system performance when nodes are loaded with high-rate beacons.

\subsection{Resilient Event-driven Message Dissemination}

Our scheme facilitates event-driven messages dissemination and reception leveraging safety beacons. We evaluate with two example scenarios: misbehavior evidence and \ac{DENM}. In the first example, every time a new malicious \ac{PC} is detected with a solid evidence (i.e., a bogus beacon and a malicious validating beacon, both properly signed; detected either locally or based on a received evidence), then the evidence is disseminated to neighbors. In the second example, nodes disseminate more general event-driven messages, i.e., \acp{DENM}, at a given rate along with misbehavior evidences, and we evaluate with two event validation metrics: event waiting time and event acceptance ratio. In this evaluation, we consider an average \ac{DENM} generation interval of $30s$ that follows exponential distribution by each node. The first of $\beta_2$ repetitions is disseminated $20ms$ after the one-time key disclosure, and the subsequent $\beta_2 - 1$ repetitions are disseminated with an interval of $20ms$ from the previously one.

\begin{figure*}[t]
	\centering
	\begin{subfigure}[b]{.33\textwidth}
		\includegraphics[width=\columnwidth]{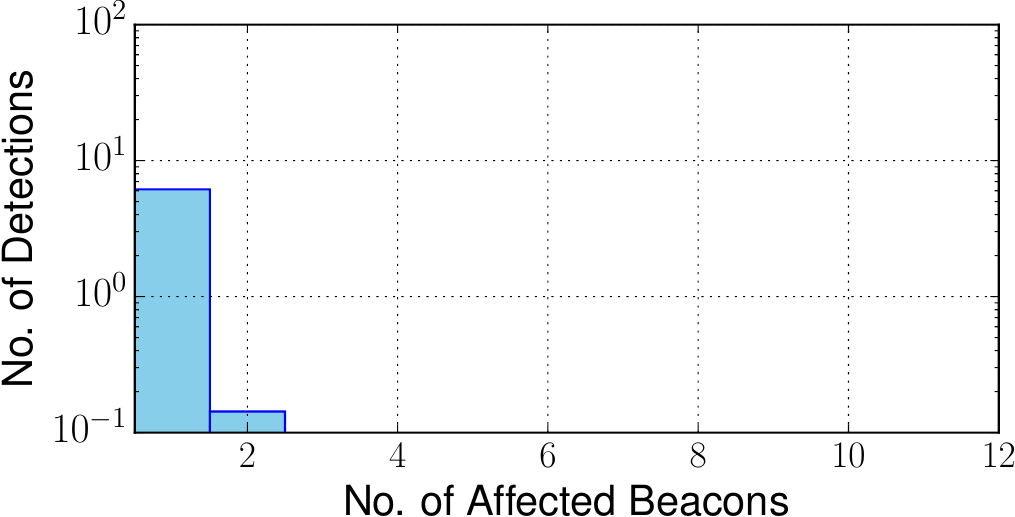}
		\caption{}
		\label{subfig_affect_advRatio_0.1}
	\end{subfigure}\hfill
	\begin{subfigure}[b]{.33\textwidth}
		\includegraphics[width=\columnwidth]{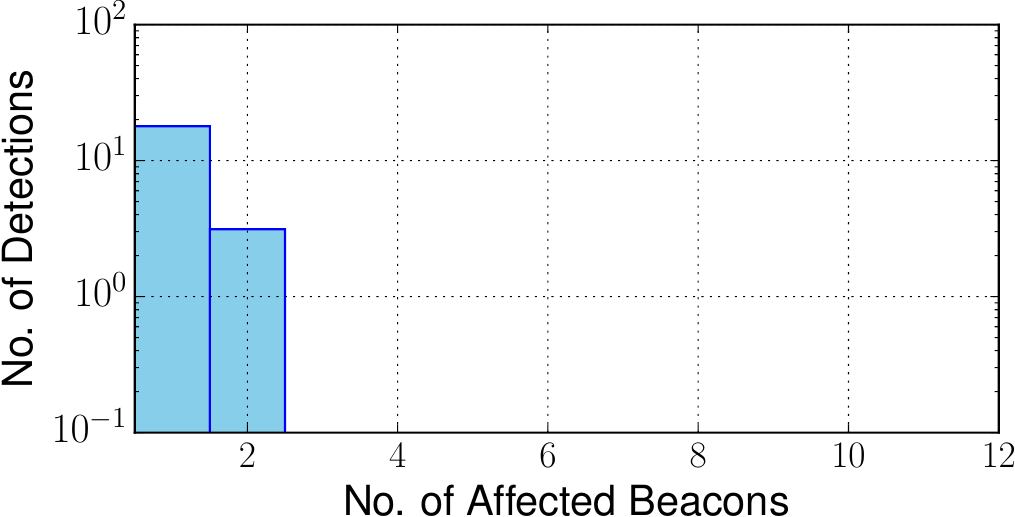}
		\caption{}
		\label{subfig_affect_advRatio_0.3}
	\end{subfigure}\hfill
	\begin{subfigure}[b]{.33\textwidth}
		\includegraphics[width=\columnwidth]{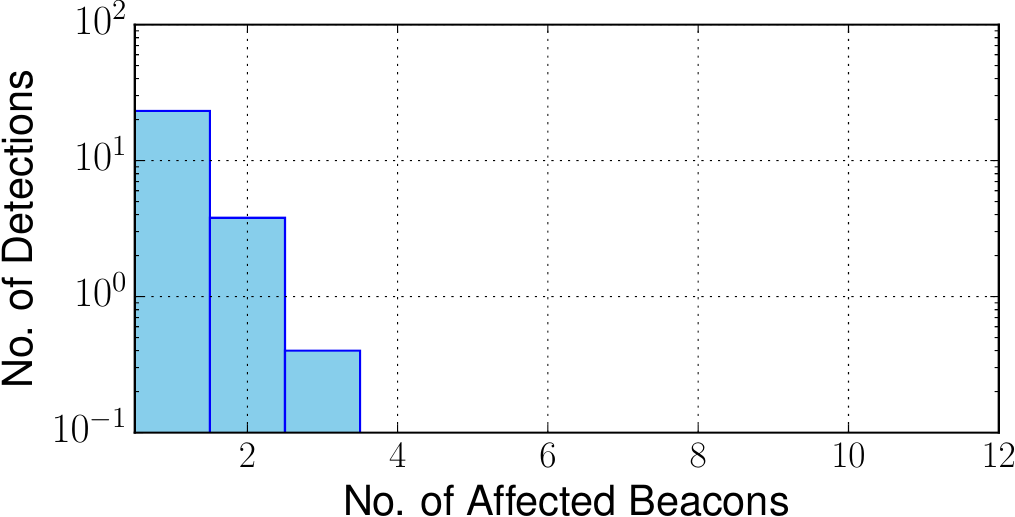}
		\caption{}
		\label{subfig_affect_advCheckProb_0.2}
	\end{subfigure}\hfill
	\caption{Histogram of numbers of affected beacons with full protection and misbehavior evidence dissemination under the DoS attack and the presence of malicious nodes. $Ratio_{adv} = 0.1, 0.3, 0.5$.}
	\label{fig_affect_advRatio}
	
\end{figure*}

We start with the first example: misbehavior evidence dissemination.
In our simulation, misbehavior evidence facilitator occupies one of $\alpha$ positions, so that $\alpha - 1$ cooperative verifiers are piggybacked when a misbehavior evidence needs to be disseminated.
Moreover, if a misbehavior evidence dissemination is ongoing, another new misbehavior evidence will be queued until the $\beta_1 * \beta_2$ repetitions conclude.
We are especially concerned with the effectiveness of this new component on malicious node detection. \Cref{fig_affect_advRatio} shows affected beacon amounts when both probabilistic checking and cross-checking mechanisms are applied.
We see the results improve again (over \cref{subfig_affect_advNoSpreadCheckProb_0.2}) thanks to proactive malicious node eviction based on the misbehavior evidences.
With more realistic $Ratio_{adv}$ values (0.1 and 0.3 in \cref{subfig_affect_advRatio_0.1,subfig_affect_advRatio_0.3}), our scheme effectively thwarts the malicious nodes and significantly minimizes the vulnerability window introduced by cooperative verification.
Our scheme still guarantees timely verification and acceptance of benign beacons, while the above detection features protect benign nodes from being attacked by the malicious nodes and DoS attacks.

For the second example, we are concerned with event waiting time and event acceptance ratio.
In this scenario, each beacon could carry at most one misbehavior evidence facilitator and at most one \ac{DENM} facilitator; both share $\alpha$ positions with cooperative verifiers.
\emph{Event waiting time} denotes delay between event generation at sender and successful event verification at receiver, and \emph{event acceptance ratio} denotes ratio of received distinct event messages that match locally cached event facilitators out of total received distinct event messages.
A received valid event message, without a local cached matching facilitator, will be dropped immediately, because it is indistinguishable from bogus ones.
Here, we consider a type of DENM disseminated for collision avoidance~\cite{c2c-irc}.
Once a \ac{DENM} is triggered, it is repeated three times with an interval of 100 ms, thus $\beta_1 = 1$ and $\beta_2 = 3$ by default.
We consider a \ac{DENM} lifetime of 2s, according to the standard~\cite{c2c-irc}.
We assume \acp{DENM} are assigned higher priority than safety beacons.
Whenever \acp{DENM} are received, they are verified before verifying queued safety beacons.

\begin{figure*}[t]
	\centering
	\begin{subfigure}[b]{.24\textwidth}
		\includegraphics[width=\columnwidth]{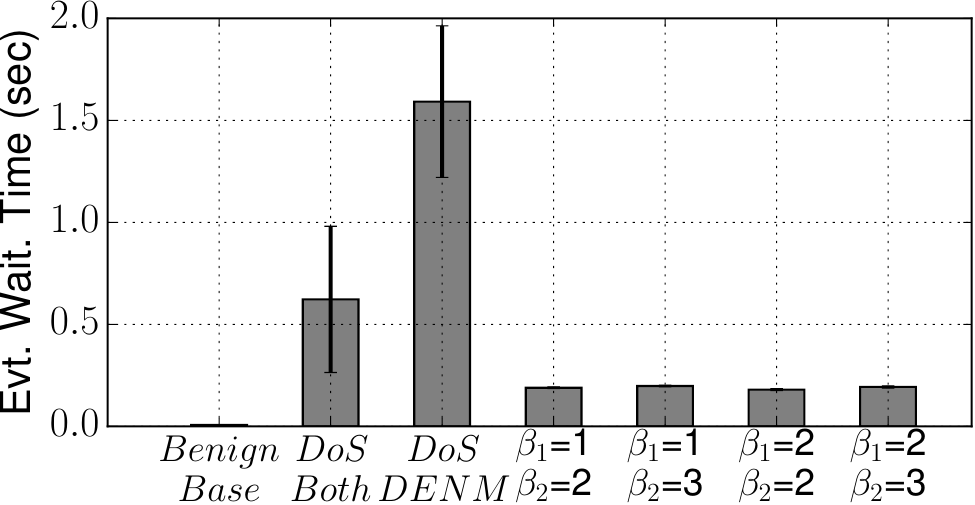}
		\caption{}
		\label{subfig_dos_eventWait_event}
	\end{subfigure}\hfill
	\begin{subfigure}[b]{.24\textwidth}
		\includegraphics[width=\columnwidth]{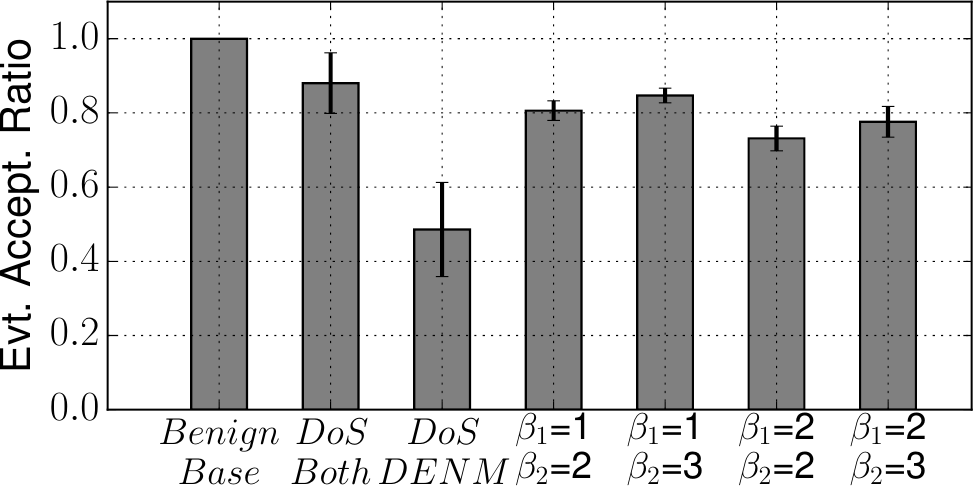}
		\caption{}
		\label{subfig_dos_eventExpire_event}
	\end{subfigure}\hfill
	\begin{subfigure}[b]{.24\textwidth}
		\includegraphics[width=\columnwidth]{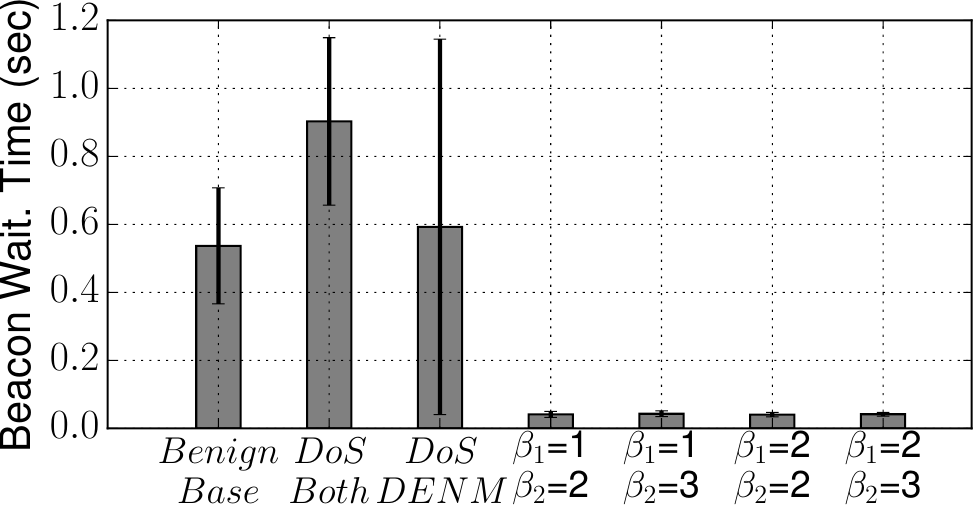}
		\caption{}
		\label{subfig_dos_wait_event}
	\end{subfigure}\hfill
	\begin{subfigure}[b]{.24\textwidth}
		\includegraphics[width=\columnwidth]{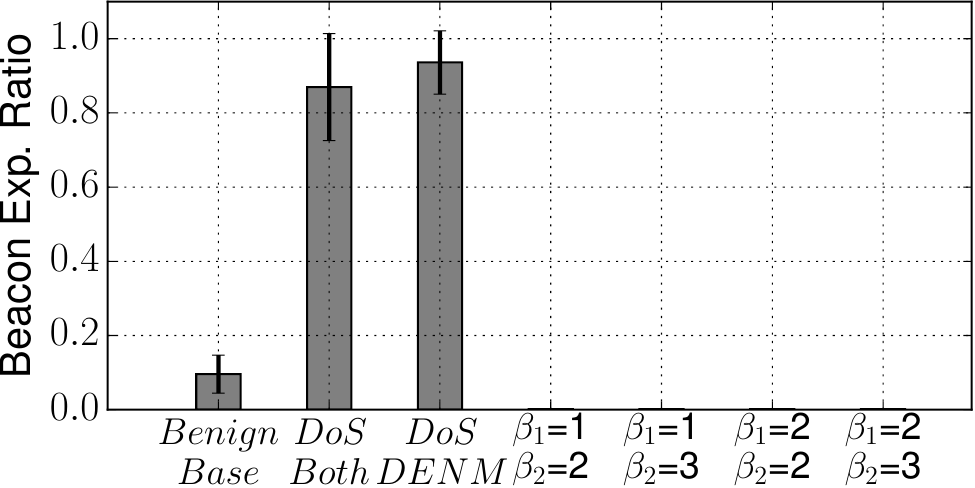}
		\caption{}
		\label{subfig_dos_expire_event}
	\end{subfigure}\hfill
	\caption{Event and beacon validation metrics with full protection under the DoS attack.}
	\label{fig_event_dos_full}
\end{figure*}

We evaluate the two metrics for the baseline scheme and our scheme with different $\{\beta_1, \beta_2\}$ combinations (\cref{subfig_dos_eventWait_event,subfig_dos_eventExpire_event}).
For the baseline scheme, event acceptance ratio is calculated with $(1 - expiration\ ratio)$.
In the benign network, the baseline scheme provides a low waiting time with almost 100\% \ac{DENM} acceptance, due to low \ac{DENM} arrival rate and higher priority given to the \acp{DENM}.
We consider two \ac{DoS} attack scenarios: the attackers flood with both beacons and \acp{DENM}, or \acp{DENM} only.
In the former scenario, each attacker broadcasts beacons and \acp{DENM} both at 125Hz respectively (thus, 250Hz in total).
In the latter scenario, each attacker broadcasts only \acp{DENM} at 250Hz.
As expected, event waiting time for the former scenario is around 0.6s, lower than around 1.6s for the latter scenario.
At the same time, the former scenario exhibits an event acceptance ratio of around 90\%, compared to around 50\% for the latter scenario.
With our scheme, average event waiting times for all combinations are around 0.17s, and around 80\% of received event messages are kept and verified.
Moreover, it provides reasonably low beacon waiting time and expiration ratio (\cref{subfig_dos_wait_event,subfig_dos_expire_event}): a significant improvement from the baseline scheme.
Beacon waiting time for the \ac{DoS}-with-\ac{DENM}-only scenario shows a large confidence interval, due to insufficient sample size collected for beacon waiting time, given high beacon expiration ratio (\cref{subfig_dos_expire_event}).
For example, in one of the seeded simulation runs, only one sample for beacon waiting time was recorded with very low value, while all the rest are expired.

\subsection{Discussion}

Our scheme is resilient to \ac{DoS} attacks and reliable in an highly-loaded benign network.
However, in lightly-loaded benign networks (see \cref{subfig_dos_eventWait_event,subfig_dos_eventExpire_event} for example), the baseline scheme performs better: low waiting time and low expiration ratio, rendering the extra communication overhead and the semi-randomized \ac{LCFS} queue unnecessary.
To achieve optimal performance in all conditions, our future work will include dynamic switching from the baseline to our scheme, based on real-time network conditions.

We provide a preliminary evaluation for one type of \ac{DENM}, covering two important metrics - message waiting time and acceptance ratio.
However, considering the existence of various \acp{DENM} and event-driven messages, an extended evaluation for the applicability of our scheme is required, especially addressing functional requirements of various safety applications.

Our simulations are built on standard-compliant IEEE 802.11p for \ac{V2V} communication.
As an evolution of IEEE 802.11p, the proposed IEEE 802.11bd provides a potential maximum rate of 87.75 $Mbps$~\cite{triwinarko2021phy}, much higher than the maximum data rate of 27 $Mbps$ for IEEE 802.11p.
Moreover, the current 5G and upcoming 6G for C-\ac{V2X}~\cite{wang20206g} could also coexist with IEEE 802.11p/bd.
Continuously developing communication standards together with relatively constant cryptographic delays (for sustainable security, as discussed in~\cref{sec:related}) would aggravate clogging \ac{DoS} attacks.
The higher the bandwidth provided by upcoming advanced communication technologies (thus, the higher maximum message rate an attacker has at her disposal), the more relevant a scheme as this one is: high bogus message rates can overwhelm even high-end CPU provisioning.

%% file: section_compsec/conclusion.tex
\section{Conclusions}
\label{sec:conclusion}

We propose a scalable, \ac{DoS}-resilient safety message verification scheme, provides vehicles with efficient message verification and protection against \ac{DoS} attacks, orthogonal to the underlying physical layer communication technology.
Simulation results show average beacon verification latency of $50 ms$ with less than 1\% expiration ratio even under \ac{DoS} attacks.
Our scheme minimizes vulnerability to misuse of cooperative verification, thanks to probabilistic checking and verifier cross-checking.
It also facilitates reception and verification of different types of \ac{V2V} messages leveraging safety beacon format extension.
With a realistic \ac{DENM} dissemination model, simulation results show 80\% of message acceptance ratio with an average latency less than $200 ms$, compared to 50\% - 100\% message expiration ratio with the baseline scheme.